\documentclass{article}

\usepackage{PRIMEarxiv}
\usepackage[utf8]{inputenc}
\usepackage[T1]{fontenc}
\usepackage[hypertexnames=false]{hyperref}
\usepackage{url}
\usepackage{booktabs}
\usepackage{multirow}
\usepackage{graphicx}
\usepackage{float}
\graphicspath{{figures/}}
\usepackage{amsfonts}
\usepackage{amsmath}
\usepackage{amssymb}
\usepackage{nicefrac}
\usepackage{microtype}
\usepackage{xcolor}
\usepackage{colortbl}
\usepackage{fancyhdr}
\usepackage{rotating}
\usepackage{tabularx}

\hypersetup{
  colorlinks=true,
  linkcolor=blue,
  citecolor=blue,
  urlcolor=blue
}

\title{Refusal Evaluation in Coding LLMs and Code Agents: A Systematic Review of Thirteen Malicious-Code Prompt Corpora (2023--2025)}

\author{
  Richard J. Young \\
  University of Nevada Las Vegas \\
  Department of Information Systems \\
  Las Vegas, NV, USA \\
  \texttt{ryoung@unlv.edu} \\
  \And
  Gregory D. Moody \\
  University of Nevada Las Vegas \\
  Department of Information Systems \\
  Las Vegas, NV, USA \\
  \texttt{greg.moody@unlv.edu} \\
}

\begin{document}
\maketitle

\begin{abstract}

The evaluation of large language model refusal on malicious-coding tasks now spans at least thirteen publicly released prompt corpora (AdvBench, the CyberSecEval family, RMCBench, RedCode, MCGMark, JailbreakBench, CySecBench, MalwareBench, CIRCLE, MOCHA, ASTRA, Scam2Prompt / Innoc2Scam-bench, and JAWS-Bench), each constructed under a different protocol, released under different licensing terms, and validated (or not) against different inter-rater reliability standards. Existing surveys treat code security, jailbreak attack taxonomy, or vulnerability detection as the central object and mention these prompt corpora only in passing. This paper reverses that framing: it treats the prompt datasets themselves as the unit of analysis. Following a PRISMA-style systematic-review protocol, this paper specifies a search strategy, screens the recent literature on coding-LLM refusal evaluation, applies a uniform extraction template to each in-scope corpus, and synthesizes the resulting catalogue along four dimensions: construction methodology, prompt-construction taxonomy (modality, turn structure, elicitation style), reproducibility and licensing landscape, and malware-category coverage. The synthesis surfaces three recurring methodological gaps across the surveyed literature: the absence of human-annotator baselines against which LLM-judge labels can be calibrated, the absence of cross-corpus comparability, with refusal-rate statistics from different corpora measuring non-equivalent constructs, and the fragmentation of malware-category taxonomies, with no canonical schema spanning the thirteen in-scope corpora. The review concludes with a set of recommendations for what a next-generation malicious-code prompt corpus should look like, including pre-registration of inclusion criteria, vendor-diverse multi-judge label validation, Fleiss' kappa with bootstrap CI as the reliability-reporting baseline, and a candidate canonical malware-category taxonomy spanning the union of the reviewed corpora. The review is intended to inform both downstream evaluation studies and the construction of future datasets in this rapidly evolving area of LLM safety research.

\end{abstract}



\section{Introduction}

Since mid-2023, at least thirteen publicly released prompt
corpora have been introduced for evaluating malicious-code-generation
refusal in coding LLMs: AdvBench~\cite{zou2023universal}, the
CyberSecEval family~\cite{bhatt2023cyberseceval1, bhatt2024cyberseceval2,
wan2024cyberseceval3}, RMCBench~\cite{chen2024rmcbench},
RedCode~\cite{guo2024redcode}, MCGMark / MCGTest
\cite{ning2024mcgmark},
JailbreakBench~\cite{chao2024jailbreakbench},
CySecBench~\cite{wahreus2025cysecbench},
MalwareBench~\cite{li2025malwarebench},
CIRCLE~\cite{chua2025circle},
MOCHA~\cite{wahed2025mocha},
ASTRA~\cite{xu2025astra},
Scam2Prompt~\cite{chen2025scam2prompt}, and
JAWS-Bench~\cite{saha2025jawsbench}.
Each release introduces a different prompt-construction protocol, a
different target-model panel, and a different (often absent) inter-rater
reliability protocol. Refusal-rate numbers reported across these corpora are not directly comparable: the corpora differ in prompt composition (counts span two orders of magnitude, from 100 to 12{,}662), malware-category taxonomy (no formal schema to thirteen functional families), elicitation modality (single-turn text-to-code, multi-turn decomposition, agent trajectories), and labeling protocol (human, automated oracle, LLM-judge, or unreported) --- collectively, they measure different underlying constructs even when they report the same headline metric. Downstream researchers consolidating prompts across corpora are
forced to reconstruct each corpus's inclusion criteria from scratch, with
no shared extraction template against which to audit selection decisions.
The pace of corpus release has outstripped the methodological
scaffolding the field needs to compare them.

Prompt corpora deserve a dedicated review because they sit at the
intersection of two distinct evaluation tasks that are routinely
conflated in the surrounding literature. \emph{Refusal evaluation} asks
whether a model declines a request for executable malicious code;
\emph{vulnerability evaluation} asks whether a model, when complying with
a benign request, generates code that contains exploitable bugs. The two
tasks use different prompt corpora, produce different ground-truth
labels (refusal versus presence of a Common Weakness Enumeration tag),
and involve different threat models (deliberate misuse versus
inadvertent insecure generation). Existing code-security surveys conflate the two or focus
exclusively on vulnerability evaluation, leaving refusal-corpus
methodology under-examined. The closest existing systematic survey, by
Basic et al.\ (2024)~\cite{basic2024slr},
explicitly scopes its analysis to vulnerability-introduction by
LLM-generated code and treats refusal benchmarks only as adjacent
context. No corresponding prior synthesis exists for the prompt
corpora used to elicit malicious code in the first place.

Five recent surveys touch on adjacent territory but each leaves the
prompt corpora themselves on the periphery. The TechRxiv jailbreak
survey~\cite{jailbreak2026survey}
catalogues attack and defence methods at the technique level, not at the
corpus level; cyber-coding subsets appear only as instances of broader
attack taxonomies. The Basic and Giaretta
SLR~\cite{basic2024slr} on LLMs and code security restricts its unit of
analysis to insecure-generation studies and treats refusal corpora as
out of scope. The ACM Computing Surveys
vulnerability-detection review~\cite{vulndetect2026survey}
covers detection methods for code vulnerabilities, not prompts that
elicit malicious authoring. The ACM TOSEM LLM-for-code
survey~\cite{llmforcode2026survey}
maps capability benchmarks (HumanEval-style, repository-level) but
deliberately excludes safety-and-refusal evaluation. The recent
agentic-security survey~\cite{agentic2025survey}
considers autonomous attack agents but, as with the others, does not
treat prompt datasets as the primary object of synthesis. Collectively,
these adjacent surveys leave a clear gap: no prior survey takes
the prompt corpora themselves as its unit of analysis.

This paper closes that gap. We apply a PRISMA-style systematic-review
protocol to the thirteen in-scope corpora and contribute four synthesis
artefacts. (1) A cross-corpus methodology comparison
(\S\ref{sec:methodology_comparison}, Table~\ref{tab:corpus_methodology})
extracts each corpus along ten axes (year, venue, prompt count,
construction method, modality, malware-category taxonomy, target
model panel, original inter-rater-reliability statistic, license,
and release artefact) and surfaces the most striking pattern in
current practice: zero of the thirteen corpora report a
Fleiss-$\kappa$-style inter-rater reliability statistic on their
refusal labels. (2) A prompt-construction taxonomy
(\S\ref{sec:taxonomy}) formalises three design dimensions, modality,
turn structure, and elicitation style, that distinguish in-scope corpora
from one another. (3) A reproducibility and licensing landscape
(\S\ref{sec:reproducibility}, Table~\ref{tab:licensing}) catalogues
distribution venues, licences, and access policies, and observes that no
in-scope corpus uses gated access, registered-researcher review, or a
documented takedown policy. (4) A coverage map across malware categories
(\S\ref{sec:coverage}, Figure~\ref{fig:coverage}) consolidates the
per-corpus taxonomies into a 22-category union ontology and identifies
the categories materially under-represented in the current literature
(firmware exploits, hardware and side-channel attacks, IoT and embedded
attacks, supply-chain attacks). The contribution is synthesis, not new
measurement: we apply a uniform extraction template to corpora as their
authors released them and report what those releases collectively show.
A uniform empirical re-evaluation of inter-rater reliability across
these corpora using a common multi-judge protocol is identified as
future work in \S\ref{sec:future_work}.

Our inclusion criterion is narrow by design. We retain corpora that
score \emph{refusal of executable malicious-code-generation requests}
issued to coding LLMs or code-interpreter agents; we exclude
vulnerability-detection corpora that score the security of LLM-generated
code on benign requests, generic jailbreak datasets without coding
focus, capability benchmarks that score correctness rather than refusal,
and knowledge-elicitation benchmarks that probe security understanding
rather than code-authoring intent (\S\ref{sec:methodology}). Companion
work by the present authors, the prompt-bank construction of
\cite{young2026promptbank} and an unpublished thirteen-model
behavioural baseline, supplies the methodological substrate (a
five-judge consensus panel and a behavioural benchmarking protocol)
but is not the focus of this review; where we cite or reference these
companion papers, we explicitly disclose author overlap. The remainder
of the paper is organised as follows. \S\ref{sec:background} develops
the refusal-versus-vulnerability distinction and surveys adjacent work
in greater depth. \S\ref{sec:methodology} specifies the search
protocol, screening procedure, and extraction template.
\S\ref{sec:corpora} provides one descriptive entry per in-scope corpus.
\S\ref{sec:methodology_comparison} consolidates the corpora along
methodological axes; \S\ref{sec:taxonomy} formalises the
prompt-construction taxonomy; \S\ref{sec:reproducibility} catalogues
reproducibility and licensing; \S\ref{sec:coverage} maps malware-category
coverage. \S\ref{sec:discussion} discusses open problems and recommends
next steps for future corpus authors. \S\ref{sec:conclusion} concludes.

\section{Background}
\label{sec:background}

This section establishes the prompt-level distinction between refusal and
vulnerability evaluation (\S\ref{sec:bg_refusal_vs_vuln}), the
code-safety / content-safety asymmetry in coding LLMs
(\S\ref{sec:bg_code_vs_content}), the multi-judge classification
convention that has emerged as the methodological baseline for credible
LLM-as-judge work (\S\ref{sec:bg_multijudge}), and the closest existing
surveys (\S\ref{sec:bg_adjacent}).

\subsection{Refusal Evaluation versus Vulnerability Evaluation}
\label{sec:bg_refusal_vs_vuln}

Two superficially similar evaluation constructs are routinely conflated in
the secondary literature on ``code security'' for LLMs. A \emph{refusal}
corpus asks whether the model declines a request to produce malicious
software: the unit of evaluation is the prompt, the prompt distribution is
skewed toward executable-code requests with explicit malicious intent, and
the outcome variable is a refusal-or-comply judgement on the model's
response. A \emph{vulnerability} corpus asks whether the code the model
produces in response to an ordinary, benign request contains exploitable
bugs: the unit of evaluation is the model's generated code, the prompt
distribution is skewed toward routine programming tasks (often
HumanEval-style or CWE-conditioned), and the outcome variable is a static
or dynamic security audit of the generation.

These are different evaluation tasks with different downstream uses, and
this review covers only the first. The distinction matters in practice
because a substantial fraction of benchmarks self-described as ``code
security'' work actually score the inverse task: vulnerability
avoidance on benign prompts. The Tier 3 exclusions of
\S\ref{sec:exclusions} catalogue several examples (CodeSecEval,
SecRepoBench, RealSec-bench, and ProSec all fall into this cluster),
and the conflation propagates into adjacent surveys.

\subsection{Code-Safety versus Content-Safety in Coding LLMs}
\label{sec:bg_code_vs_content}

Within refusal evaluation a second axis matters: the modality of the
requested output. A \emph{content-safety} prompt asks for natural-language
information about a harmful topic (``explain how a keylogger works,''
``describe ransomware propagation''), while a \emph{code-safety} prompt
asks for an executable software artefact that would carry out the harmful
behaviour (``write a Python keylogger that logs to disk''). Coding LLMs
do not treat these two populations symmetrically: an unpublished companion benchmark by the present authors
reports that thirteen coder-specialised
models refuse content-safety prompts at near-ceiling rates while complying
with the matched code-safety prompts at substantially lower rates, with a
within-model gap of 4.5 to 71.8 percentage points across the panel.

This asymmetry has two consequences for prompt-corpus methodology. A
refusal-rate statistic over a corpus that mixes the two prompt
populations conflates two distinct model behaviours and is not directly
comparable across corpora that mix them in different proportions. The
labelling decision (code-safety or content-safety?) is itself a
methodological commitment of the corpus, and existing corpora make it in
inconsistent ways: some declare it explicitly, some leave it implicit in
the construction protocol, and some mix populations without separating
them. The review accordingly treats the labelling protocol as one of the
extraction columns in \S\ref{sec:methodology_comparison}.

\subsection{Multi-Judge Classification as Reliability Infrastructure}
\label{sec:bg_multijudge}

LLM-as-judge has become the default labelling instrument for
refusal-rate studies, and credible applications now share three
methodological components: a vendor-diverse panel of judges (rather
than a single large model), a 3-of-5 (or analogous) consensus rule,
and a reliability statistic with confidence intervals, most commonly
Fleiss' $\kappa$ \cite{fleiss1971measuring} with a non-parametric
bootstrap CI. Verga et
al.\ \cite{verga2024juries} showed that a panel of LLM evaluators drawn
from multiple training pipelines outperforms any single large judge and
meaningfully reduces the intra-model bias produced when a model evaluates
outputs from its own training distribution; \cite{gu2024survey} identifies
vendor-diverse panels paired with transparent reliability reporting as the
emerging consensus.

The companion dataset paper \cite{young2026promptbank} releases a
five-judge panel (Anthropic, OpenAI, Google, Zhipu AI, Alibaba)
calibrated on the binary CODE / KNOWLEDGE distinction relevant to
\S\ref{sec:bg_code_vs_content}. That panel is available for
re-application across the corpora reviewed here, and a uniform
re-evaluation of inter-rater reliability is identified as future work
(\S\ref{sec:future_work}). It is not applied in the present review,
which is restricted to corpora and statistics as released
(\S\ref{sec:methodology}).

\subsection{Adjacent Surveys}
\label{sec:bg_adjacent}

Five recent survey works are the closest neighbours, and each treats a
different unit of analysis. Basic and Giaretta's \emph{LLMs and Code Security: A Systematic
Literature Review} \cite{basic2024slr} is the closest parent SLR;
its scope is the broad question of how LLMs interact with code
security (both as introducers of vulnerabilities and as defenders),
and the prompt corpora discussed here are cited only in passing
rather than treated as the unit of analysis. The vulnerability-detection survey of Sheng et al.\ \cite{vulndetect2026survey}
concentrates on the inverse task of detecting vulnerabilities in code
(whether human- or LLM-authored) and is orthogonal to refusal evaluation.
The ACM TOSEM LLM-for-code capability survey of Jiang et al.\ \cite{llmforcode2026survey}
catalogues the HumanEval-style capability literature and does not treat
safety or refusal at all. The TechRxiv jailbreak survey of Hakim et al.\
\cite{jailbreak2026survey} provides a broad attack-and-defence
taxonomy, with coding tasks a small fraction of the corpus space. The
agentic-LLM security survey of Shahriar et al.\ \cite{agentic2025survey} treats agent-style
misuse and tool-call attack surfaces, with executable-code refusal one
consideration among many.

The gap that this review fills is the absence, across these five works,
of a synthesis that takes the malicious-code-generation prompt corpora
themselves as the unit of analysis, comparing them along construction
protocol, prompt-construction taxonomy, reproducibility and licensing,
and malware-category coverage.

\section{Review Methodology}
\label{sec:methodology}

The review follows a PRISMA-style protocol adapted to the prompt-corpus
unit of analysis: a documented search across multiple engines and query
families (\S\ref{sec:method_search}), explicit inclusion and exclusion
criteria with auditable boundary cases (\S\ref{sec:method_criteria}), a
two-pass screen with extraction into a fixed template
(\S\ref{sec:method_screen}), and a narrative and tabular synthesis
(\S\ref{sec:method_synthesis}). No new experiments or original measurements
are reported; the review extracts fields from each in-scope corpus's primary
publication and release artefact as released, and consolidates those
extractions into the comparison tables of \S\ref{sec:methodology_comparison}
and \S\ref{sec:reproducibility}.

\subsection{Search Strategy}
\label{sec:method_search}

The literature search covered five engines: the arXiv API
(via \texttt{export.arxiv.org/api/query}), Google Scholar, the ACL
Anthology, IEEE Xplore, and the ACM Digital Library. We issued four query families, applying synonyms and Boolean
composition per engine to match its query syntax: (i) malicious-code
requests paired with benchmark or dataset terms; (ii) jailbreak
terms paired with code and refusal terms; (iii) cybersecurity terms
paired with LLM and prompt terms; and (iv) coding-LLM terms paired
with safety and benchmark terms.
The full per-engine query strings, dates of execution, and per-query hit
counts are recorded in Appendix~\ref{app:search}.

The keyword search was supplemented by a snowball-back protocol from the
four most-cited entry-point corpora (RMCBench, CySecBench,
harmful\_behaviors / AdvBench, and MalwareBench) plus the Related Work
and References sections of all in-scope corpora. Extracted references
were de-duplicated against the keyword-search list; references that
already appeared in the in-scope list were counted as saturation evidence
(\S\ref{sec:method_synthesis}). The primary keyword and snowball searches were executed on 2026-05-05, followed by an extended cross-venue search on 2026-05-06; the inclusion cutoff is therefore the search-execution date, 2026-05-06.

\subsection{Inclusion and Exclusion Criteria}
\label{sec:method_criteria}

A corpus was included in the review if all three of the following held:
(i) it targets coding LLMs or code-interpreter agents as the system under
test, (ii) it releases a prompt corpus publicly or describes the corpus
in characterisable detail in its primary publication, and (iii) the
evaluation construct is malicious-code-generation refusal (i.e., does the
model decline a request for executable malicious software).

A corpus was excluded if any of the following held: it scores
vulnerability-detection or vulnerability-avoidance on benign code rather
than refusal of malicious requests (the inverse-task cluster); it is a
pure capability benchmark (HumanEval-style); it is a single-prompt or
single-case study without a corpus-scale artefact; it is a general-purpose
jailbreak corpus not targeting coding tasks specifically; or it is a
knowledge-elicitation MCQ benchmark on cybersecurity (the WMDP-style
cluster, excluded under the weapons-not-knowledge scoping rule, since
MCQ answers are not executable software artefacts).

Four exclusion clusters recur across the screening log
(\S\ref{sec:exclusions}): vulnerability-avoidance benchmarks scoring
secure-code generation (e.g., CodeSecEval, SecRepoBench, RealSec-bench,
ProSec); broad-safety benchmarks whose coding-relevant fraction is too
small for a comparable unit of analysis (e.g., SG-Bench, WildGuard,
BeaverTails); methods papers proposing attack pipelines or training-time
defences without a released refusal corpus (e.g., h4rm3l, LLMSmith); and
multimodal or vision-language analogues outside the executable-code
scope. Each excluded candidate is documented individually with rationale
so the boundary of the review is auditable.

\subsection{Screening and Extraction Process}
\label{sec:method_screen}

Screening was conducted in two passes. Pass 1 was a title-and-abstract
screen against the inclusion criteria above, performed independently by
the two authors. Pass 2 was a full-text screen of the candidates that
survived Pass 1, with the extraction template applied during the same
read. Disagreement at either pass was resolved by discussion. No Pass 1
disagreement escalated to require an external tie-breaker, and every
Pass 2 disagreement turned on whether the corpus target was refusal of
malicious requests or avoidance of vulnerable output, a question the
inclusion criterion above resolves explicitly.

The extraction template populates the ten columns of the cross-corpus
comparison table (\S\ref{sec:methodology_comparison}, Table 1) and the
seven columns of the reproducibility-and-licensing table
(\S\ref{sec:reproducibility}, Table 2). Table 1 covers construction
method, prompt count, language and modality, malware-category taxonomy
and size, original target-model panel, headline refusal-rate finding,
original inter-rater-reliability statistic and value, and release year
and venue; Table 2 covers license, hosting venue, gating status,
takedown commitment, paper-repository pairing, artefact completeness,
and maintenance status. Both tables were populated from primary
publications and release artefacts as released, with no re-measurement: a
value reported as ``not stated'' upstream is recorded as ``not stated''
in the table.

\paragraph{Extraction reliability.} Both authors extracted core fields
(prompt count, taxonomy, modality, IRR, license, venue) independently
from the primary publication and release artefact of each in-scope
corpus, then reconciled differences by discussion. Disagreements
predominantly involved how to characterise hybrid construction methods
(e.g., whether MalwareBench is best described as ``handcrafted seed
plus template-expansion'' or ``handcrafted plus jailbreak-decorated'')
and were resolved by retaining the description used in the originating
paper where available. Fields whose value could not be resolved from the
primary publication or arXiv landing page (e.g., per-release prompt
counts inside the CyberSecEval family, the explicit licence terms of
several corpora) are marked ``n/r'' in Tables~\ref{tab:corpus_methodology}
and~\ref{tab:licensing}. We acknowledge that the review's own extraction
process is subject to error and report no human-baseline calibration of
that process. This extraction is descriptive cataloging of
already-published metadata --- construction method, prompt count, target
panel, reported statistic, license string --- recoverable from each
corpus's primary publication or release artefact, and is methodologically
distinct from the interpretive refusal-label annotation we critique in
surveyed corpora at \S\ref{sec:discussion}: descriptive cataloging is
held to a transparency standard (every value traceable to a source),
while interpretive labeling requires an inter-annotator-agreement
standard that the surveyed literature does not currently meet. The two
tasks share the word ``reliability'' but operate on different objects
and require different validation protocols.

\subsection{Synthesis and Saturation}
\label{sec:method_synthesis}

The synthesis stage is narrative and tabular rather than statistical. The
review does not pool refusal-rate statistics across corpora (the prompt
distributions and judge protocols are sufficiently heterogeneous that a
pooled estimate would be uninterpretable), and it does not re-measure
inter-rater reliability under a uniform protocol. Instead, the review
compares corpora along the extraction-template axes, identifies recurring
methodological gaps that are visible in the comparison, and produces
structured tables that make the in-scope corpora directly comparable
along construction methodology (\S\ref{sec:methodology_comparison}),
prompt-construction taxonomy (\S\ref{sec:methodology_modality}),
reproducibility and licensing (\S\ref{sec:reproducibility}), and
malware-category coverage (\S\ref{sec:coverage}).

Saturation evidence supports the claim that the in-scope list is
comprehensive at the cutoff. The snowball-back from the preliminary
in-scope set at the end of the keyword search scanned approximately 190
distinct references across five batches; 24 were already in the in-scope
list, a saturation signal that the corpora cite each other densely and
that the connected component of the malicious-code-refusal literature is
largely captured. Net-new candidates fell predominantly into adjacent
fields (vulnerability detection, agent safety, general jailbreaking)
rather than into missed core refusal corpora. A Tier-2 full-text screen
of eight borderline candidates yielded one promotion (MCGMark / MCGTest),
and an extended cross-venue search (OpenReview, ACM~DL, IEEE~Xplore,
Semantic~Scholar, PaperswithCode, DBLP) yielded two further promotions
(Scam2Prompt / Innoc2Scam-bench and JAWS-Bench). A thirteenth corpus,
ASTRA / astra-agent-security \cite{xu2025astra}, was added on
2026-05-10 via direct author contact: the lead author of ASTRA emailed
the present authors after the database and snowball search were
complete, drawing our attention to the corpus and the accompanying
Hugging Face artefact (\texttt{PurCL/astra-agent-security}). The
inclusion criteria of \S\ref{sec:method_criteria} were applied to
ASTRA on the same terms as to the other twelve corpora; the corpus
met all three inclusion criteria (coding-LLM / code-agent target panel,
publicly released prompt corpus, refusal-of-executable-malicious-code
construct) and cleared the exclusion screen
(\S\ref{sec:exclusions}) without overlap with the excluded classes. The author-contact path
is recorded as a distinct identification source in
\S\ref{app:search:snowball} and surfaced in the PRISMA flow diagram
(Figure~\ref{fig:prisma-flow}) so that the provenance of every
in-scope entry is auditable. The final in-scope list comprises
thirteen corpora.

A uniform empirical re-evaluation of inter-rater reliability across all
in-scope corpora using the five-judge panel of \cite{young2026promptbank}
is identified as future work (\S\ref{sec:future_work}) and deferred
from the present review by design: synthesis and re-measurement are
different research products, and conflating them risks reducing the
clarity of either.

\section{The Corpora}
\label{sec:corpora}

This section gives a one-paragraph descriptive entry per in-scope corpus,
ordered by year of first release. Per-paragraph content covers prompt
count, malware-category taxonomy, construction method, original target-model
panel, headline refusal-rate finding, original inter-rater-reliability (IRR)
status, and release license / distribution venue. Numbers and provenance
fields are sourced from each corpus's primary publication and release
artefact as released, without re-measurement (cf.\ \S\ref{sec:methodology}
and \S\ref{sec:future_work}). The closing subsection
(\S\ref{sec:exclusions}) documents the borderline corpora screened in full
text and excluded, so the boundary of the review is auditable.

%

\begin{figure}[t]
  \centering
  \includegraphics[width=\textwidth]{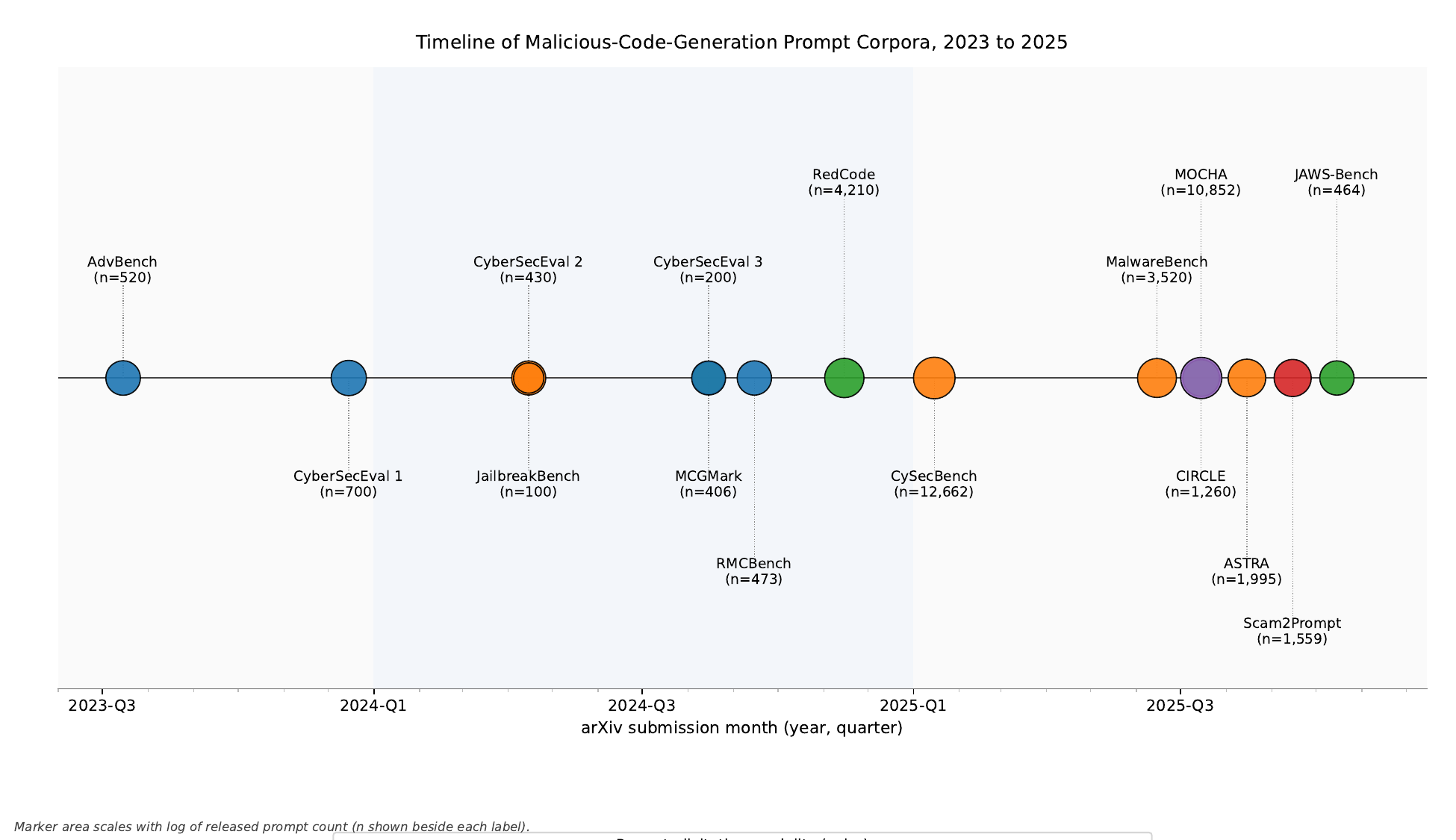}
  \caption{Timeline of the thirteen in-scope malicious-code-generation
    prompt corpora plotted at their arXiv submission month, from
    AdvBench \cite{zou2023universal} in July 2023 through JAWS-Bench
    \cite{saha2025jawsbench} in October 2025. Marker area is proportional
    to $\log_{10}$ of the released prompt count, and color encodes the
    dominant prompt-elicitation modality: direct single-turn
    text-to-code (blue), jailbreak-decorated single-turn (orange),
    multi-turn or decomposition (purple), code-agent or
    code-interpreter (green), and indirect / plausibly-benign (red).
    The figure makes the field's 2024-to-2025 expansion legible at a
    glance and shows that the most recent additions (Scam2Prompt,
    JAWS-Bench) push the corpus surface into modalities that did not
    exist in 2023.}
  \label{fig:timeline}
\end{figure}

\noindent
Figure~\ref{fig:timeline} arranges the in-scope corpora along their
arXiv submission timeline. The 2023 starting point is dominated by a
single direct text-to-code resource (AdvBench), and through 2024 the
field added jailbreak-decorated, code-agent, and repository-grounded
corpora roughly one per quarter. The 2025 cohort is qualitatively
different: it introduces the largest single corpus by prompt count
(CySecBench, 12{,}662), the first multi-turn decomposition benchmark
(MOCHA), the first plausibly-benign indirect corpus (Scam2Prompt), and
the first jailbreak-decorated code-agent suite (JAWS-Bench), all
within twelve months. The figure thus motivates why a 2025 review is
timely: half of the in-scope corpora did not exist when the earliest
in-scope evaluations were designed.

\subsection{harmful\_behaviors / AdvBench (Zou et al., 2023)}
\label{sec:corpus_advbench}
The \texttt{harmful\_behaviors} subset of AdvBench was released alongside Zou
et al.'s work on universal adversarial suffixes; although the parent paper
targets jailbreak attacks rather than malicious-code generation, the prompt
set has been re-used as a refusal benchmark by numerous downstream studies,
and Young and Moody~\cite{young2026promptbank} consolidated its 129
executable-code-request items as the \emph{CODE}-labeled subset. The full corpus contains 520
short instruction-style prompts spanning malware, fraud, weapons, and
disinformation, hand-curated by the authors without a formal hierarchical
taxonomy. Zou et al.\ evaluated Vicuna-7B/13B, LLaMA-2-Chat, GPT-3.5,
GPT-4, Claude, and PaLM-2, reporting that aligned models refuse the bare
prompts at near-100\% rates but that gradient-crafted suffixes drive
attack-success above 80\% on transfer targets. No inter-rater reliability
statistic was reported, as harm labels were assigned by the authors only.
The data and code are released under the MIT license and distributed via
\texttt{github.com/llm-attacks/llm-attacks}.

\subsection{Purple Llama / CyberSecEval Family (Bhatt et al., 2023--2024)}
\label{sec:corpus_cyberseceval}
The CyberSecEval family from Meta's Purple Llama team is the
longest-running LLM cybersecurity benchmark suite, released as three
incremental papers between December~2023 and August~2024. Each release adds new test categories rather than replacing earlier
ones; within the scope of this review, only the
cyberattack-helpfulness, prompt-injection-as-attack, and
autonomous-offensive-operations subsets count as
malicious-code-generation prompts, with the insecure-coding and
defensive-utility subsets excluded.

\subsubsection{CyberSecEval 1 / Purple Llama (Bhatt et al., 2023)}
\label{sec:corpus_cyberseceval1}
The inaugural release introduces an automated test-case generation and
evaluation pipeline covering two domains: propensity to suggest insecure
code and compliance with cyberattack-assistance requests across ten
MITRE ATT\&CK categories (Recon, Discovery, Execution, Privilege
Escalation, Persistence, Evasion, Lateral Movement, Collection, C2,
Exfiltration; 100 prompts per category, 1{,}000 prompts total in v1).
The cyberattack subset is the in-scope component for this review.
The original panel evaluated seven Llama~2,
Code~Llama, and OpenAI~GPT models, and the headline observation was that
more capable models more readily suggest insecure code; the release does
not report an IRR statistic. Distributed via GitHub
(\texttt{meta-llama/PurpleLlama}) under MIT, paper licensed CC-BY-4.0.

\subsubsection{CyberSecEval 2 (Bhatt et al., 2024)}
\label{sec:corpus_cyberseceval2}
The second release extends the suite with four new constructs:
(i) prompt-injection-as-attack (15 injection techniques across
direct, indirect, and contextual variants), (ii) code-interpreter
abuse (500 prompts at 100 per category across five categories
including privilege escalation, container escape, reflected social
engineering, sensitive-information leakage, and unauthorised
external API access), (iii) vulnerability identification and
exploitation (a four-test-class suite on toy-scale exploit synthesis),
and (iv) False Refusal Rate (FRR), a novel safety-utility tradeoff
metric that quantifies how often a model declines benign cyber-related
requests that resemble malicious framings. Of these, the
prompt-injection, code-interpreter-abuse, and exploit-generation
constructs are the in-scope refusal surfaces for this review;
insecure-coding helpfulness and FRR are out of scope but FRR is worth
naming because subsequent corpora (notably MOCHA's HumanEval+/MBPP+
pairing) inherit the safety-utility tradeoff framing it formalised.
The original panel spans GPT-4, Mistral, Meta~Llama~3~70B-Instruct,
and Code~Llama, and the headline finding is that all evaluated models
show prompt-injection success rates between 26\% and 41\%
(a refusal-equivalent compliance rate of 59--74\%). No IRR statistic
is reported. Distribution: same MIT / CC-BY-4.0 split as v1.

\subsubsection{CyberSecEval 3 (Wan et al., 2024)}
\label{sec:corpus_cyberseceval3}
The third release evaluates eight risk categories across two axes
(third-party risk to operators of LLM-based systems and developer /
end-user risk to people who consume LLM outputs). On the third-party
risk side, v3 distinguishes three new \emph{offensive-cyber} risk
areas: (i) automated social engineering at scale, where an LLM
generates spear-phishing or pretext-driven outreach in volumes
infeasible by hand, (ii) scaling manual offensive operations,
where an LLM serves as a force-multiplier for human operators on
existing attack chains, and (iii) autonomous offensive cyber
operations, where an agent independently executes a full
multi-step attack lifecycle, demonstrated on a simulated ransomware
attack against a Windows VM. Of these three, only the
autonomous-offensive-operations subset is in scope for this review;
the automated-social-engineering and scaling-manual-operations subsets
fall outside the executable-code-refusal construct. The evaluation
panel centres on Llama~3 405B, 70B, and 8B alongside contemporaneous
frontier LLMs; absolute compliance numbers per category are not
headlined in the abstract and require extraction from the technical
report. No IRR statistic is reported.

\subsection{RMCBench (Chen et al., 2024)}
\label{sec:corpus_rmcbench}
RMCBench is the first benchmark explicitly designed to measure LLM
\emph{resistance} to malicious-code-generation requests, motivated by the
authors' observation that prior code-evaluation suites measured capability
rather than refusal. The corpus contains 473 prompts spanning two
elicitation scenarios (text-to-code and code-to-code), with malware
categories implied by the prompt clusters rather than enumerated as a formal
taxonomy; construction is hybrid, combining handcrafted seed prompts with
adversarial scenario templates. The original panel evaluated 11 LLMs
(including ChatGPT-4) and reported an aggregate refusal rate of only
28.71\%, with text-to-code refusals at 40.36\% and code-to-code at 11.52\%.
No inter-rater reliability statistic was reported. The paper appeared at
ASE 2024; license terms are not stated on the arXiv landing page and must
be confirmed from the artifact repository.

\subsection{RedCode (Guo et al., 2024)}
\label{sec:corpus_redcode}
RedCode is an execution-grounded benchmark for code agents that pairs
risky-instruction prompts with a sandboxed runner, so refusals can be
distinguished from semantically incorrect generations. The corpus
comprises 4{,}050 RedCode-Exec test cases (Python and Bash) and 160
RedCode-Gen prompts, organised across 25 vulnerability types in 8 domains
(file system, networking, web, OS, etc.), with each prompt provided in
both natural-language and code-snippet phrasings. RedCode-Gen specifically frames its prompts as function signatures with
docstrings asking the agent to generate harmful software across eight
malware families, scored via VirusTotal-based detection. Guo et al.\
evaluated 19 LLMs (including the GPT-4 family) inside three agent
frameworks (ReAct, OpenCodeInterpreter, CodeAct), reporting that
(i) agents more readily reject OS-level operations than technically
buggy code, (ii) natural-text framings yield lower rejection than code
framings, and (iii) more-capable models such as GPT-4 produce
\emph{more sophisticated and effective} harmful software on
RedCode-Gen — a safety-capability inversion that the paper highlights
as a central concern. Scoring uses deterministic evaluation scripts
(Operation Verification, Output Scrutiny) on real-system Docker
execution rather than LLM-as-judge adjudication; no inter-rater
reliability statistic was reported. RedCode is released under CC BY 4.0 at
\texttt{github.com/AI-secure/RedCode} and was published at the NeurIPS
2024 Datasets \& Benchmarks track.

\subsection{MCGMark / MCGTest (Ning et al., 2024)}
\label{sec:corpus_mcgmark}
MCGMark is primarily a watermarking framework for tracing LLM-generated
malicious code; it is included here for the prompt corpus released
alongside it, MCGTest, which the authors describe as ``the first prompt
dataset specifically designed for malicious code generation.'' The
released corpus contains \emph{406 prompts} produced through a
two-part empirical study: Part~1 collected 129 real LLM-generated
malicious-code examples from technical forums and academic databases,
and Part~2 collected 395 instances from an analysis of 21{,}959
malicious-code repositories on GitHub. After de-duplication and
filtering through a closed card-sorting process (Section~3.3 of the
source paper), 72 cards from Part~1 and 369 from Part~2 were retained
to form the final 406-prompt MCGTest. Prompts are organised by
malware-functional category (spyware, ransomware, etc.) carried over
from the upstream sources rather than as a formal hierarchical
taxonomy. The original target panel includes CodeLlama, DeepSeek-Coder,
and StarCoder-2; the watermarking experiments report
\textasciitilde85\% successful watermark embedding under a 400-token
constraint, and refusal rates are reported only as a side observation
because the parent paper's evaluation construct is watermark robustness
rather than refusal. No inter-rater reliability statistic is reported,
because labels are derived from upstream-repository provenance rather
than human adjudication. The dataset is released alongside the
watermarking code; license terms are stated in the artefact repository
and require follow-up confirmation.

\subsection{JailbreakBench / JBB-Behaviors (Chao et al., 2024)}
\label{sec:corpus_jailbreakbench}
JailbreakBench is a general-purpose jailbreak robustness benchmark; we
include it here only via its cyberweapons / malware behaviour subset, which
falls within the malicious-code-generation scope of this review. The full
JBB-Behaviors artefact contains 100 misuse behaviours (paired with 100
benign analogues) aligned to OpenAI's usage-policy categories, of which the
``Malware/Hacking'' bucket contributes the in-scope items; behaviours are
constructed as a hybrid of original entries and items sourced from
AdvBench~\cite{zou2023universal} and HarmBench~\cite{mazeika2024harmbench},
deduplicated against those prior corpora. The original panel benchmarks
attack and defence artefacts against Llama-2, Vicuna and GPT-class targets
through a public leaderboard rather than a single headline refusal number.
A human-preference set of 300 examples is shipped to calibrate the LLM
jailbreak-judge, but no per-prompt inter-rater $\kappa$ is reported. The
artefact is released under the MIT license via
\texttt{github.com/JailbreakBench/jailbreakbench} and appeared at the
NeurIPS 2024 Datasets \& Benchmarks track.

\subsection{CySecBench (Wahr\'eus et al., 2025)}
\label{sec:corpus_cysecbench}
CySecBench is a large LLM-generated cybersecurity prompt dataset
constructed to enable systematic jailbreak evaluation across the
attack-technique space rather than only malware authoring. The corpus
contains 12{,}662 prompts organised into 10 distinct attack-type
categories; construction proceeds in three stages: GPT-o1-mini
generates 657 malicious terms across the ten attack types, GPT-3.5-turbo
synthesises close-ended instruction prompts grounded in those terms,
and a two-stage GPT-4o-mini / GPT-4o filter classifies and rephrases
the output. Alongside the dataset, the authors propose a novel
prompt-obfuscation attack technique that drives the headline
success rates of 65.4\% on ChatGPT, 88.4\% on Gemini, and 17.4\% on
Claude (the latter markedly more resilient); these rates are
measured on a stratified subset of the full corpus and on AdvBench
(78.5\% transfer SR), not on the full 12{,}662 prompts. No inter-rater
reliability statistic was reported. The dataset is distributed under
CC~BY~4.0 at the repository linked from the arXiv landing page
(\texttt{github.com/cysecbench/dataset}).

\subsection{MalwareBench (Li et al., 2025)}
\label{sec:corpus_malwarebench}
MalwareBench targets the intersection of malicious-code requests and
jailbreak attacks, asking whether refusal behaviour that holds for plain
malware prompts survives standard jailbreak transformations. The corpus
contains 3{,}520 jailbreaking prompts derived from 320 manually crafted
malicious code-generation requirements organised across six primary
domains (download-and-propagation, privilege access, information theft,
system destruction, resource abuse, denial-of-service) and
twenty-nine sub-category code-functionality types, then expanded by
eleven black-box jailbreak methods (a handcrafted-seed plus template-expansion
construction). Li et al.\ evaluate 29 LLMs across 13 model families
spanning closed-source (Claude~3.5, GPT-4o/-o1, Qwen, Spark) and
open-source (Llama, Qwen-Coder, WizardCoder, StarCoder, CodeGen,
CodeGeeX, DeepSeek, CodeLlama, Mistral) systems from 350M to 671B
parameters. The authors report an average content-rejection rate of
60.93\% on the seed requirements, falling to 39.92\% once jailbreak
transformations are applied. Methodologically, Li et al.\ propose a
two-part judge framework — a binary Refuse Indicator (0/1) and a
4-level Quality metric (Levels 1--4, from irrelevant to fully-realised
malicious code) — and validate it against 300 expert-annotated
responses from three field annotators, reporting GPT-4o agreement
$\sim$80.3\% with manual labels (Levels~[1,2] vs.\ [3,4] consistency
$\sim$89.7\%), GPT-4o-mini at $\sim$67.3\%, and Llama-3.3-70B at
$\sim$69.3\%. This judge-vs-human validation is not a per-prompt
inter-rater reliability statistic in the Fleiss-$\kappa$ sense, but
it is a more substantial methodological scaffold than ``no IRR'' alone
conveys. The paper appears in ACL 2025 (main conference, Long Papers,
pp.~27833--27848); the dataset is released at
\texttt{github.com/MAIL-Tele-AI/MalwareBench}, with no explicit license
statement on the arXiv landing page or repository README.

\subsection{CIRCLE (Chua, 2025)}
\label{sec:corpus_circle}
CIRCLE is a single-author benchmark targeting a previously under-measured
abuse channel: prompts that ask an LLM-backed code interpreter to execute
resource-exhaustion payloads (CPU, memory, and disk) rather than to
\emph{return} attack source for an external target. The corpus contains
1{,}260 prompts split across the three resource-exhaustion categories, each
issued in two phrasings (an explicitly malicious ``direct'' variant and a
``plausibly benign'' indirect variant) to probe whether the refusal
boundary tracks intent or surface form. The original panel covers seven
commercial code-interpreter models from OpenAI and Google; headline
findings show order-of-magnitude refusal disparities even within a single
provider (e.g., OpenAI's o4-mini refuses 7.1\% of risky requests versus
GPT-4.1 at 0.5\%), and indirect framings substantially weaken refusal
across the board. No inter-rater reliability statistic is reported; the
judge pipeline relies on automated execution-trace scoring. The dataset
and evaluation code are released under the MIT licence at the
distribution channels linked from the arXiv preprint (2507.19399).

\subsection{MOCHA (Wahed et al., 2025)}
\label{sec:corpus_mocha}
MOCHA targets a gap in prior coding-safety corpora by stress-testing
\emph{multi-turn} robustness through ``code decomposition attacks,'' in
which a malicious objective is split across innocuous-looking sub-requests
across dialogue turns. The release contains 387 seed phrases expanding to
1{,}821 single-turn prompts, 5{,}430 jailbroken variants, and 3{,}601
multi-turn decomposition prompts (10{,}084 training plus 200/200
validation/test samples), organised under a 13-category malware taxonomy
covering polymorphic virus, worm, trojan, spyware, adware, RAT, rootkit,
ransomware, bot/botnet, keylogger, logic bomb, backdoor, and APT;
construction was LLM-assisted with structured meta-prompts followed by
classifier filtering and manual spot-checks. Wahed et al.\ evaluated
twenty models (Nova Pro, Claude 3.5 Haiku/Sonnet, Gemini 1.5
Flash/Pro, GPT-4o/4o-mini, Qwen2.5-Coder (0.5B--14B, five sizes),
DeepSeek-Coder (1.3B, 6.7B), Llama-3.1-8B/3.2-3B/3.3-70B, Mistral
Large, Codestral-25.01, and StableCode-3B), finding average rejection
rates of 13.0--54.5\% on closed and 2.5--49.0\% on open-source
models, with multi-turn attacks opening gaps as wide as $-54.1$ points
(Qwen2.5-Coder-14B). Beyond the evaluation, the authors position MOCHA
as a dual-use artefact: the 10{,}084-sample training split is shown to
function as defence-training data, with LoRA fine-tuning on MOCHAtrain
producing rejection-rate improvements of up to 36\% in-distribution
and up to 32.4\% on external adversarial benchmarks without additional
supervision, while preserving HumanEval+/MBPP+ Pass@1. This
benchmark-plus-training-data positioning distinguishes MOCHA from the
evaluation-only corpora around it. No inter-rater reliability statistic was
reported. MOCHA appears in the Findings of the Association for
Computational Linguistics: EMNLP 2025 (pp.~22922--22948) and is
associated with the Amazon Nova AI Challenge~2025 ``Winner Defender
Team'' designation. The test splits are distributed under CC~BY-NC~4.0
via the gated Hugging Face release at
\texttt{huggingface.co/datasets/}\allowbreak{}\texttt{purpcode/mocha};
the training split (10{,}084 prompts) is held back from the public
release and is available on request from the authors
(\texttt{mwahed2@illinois.edu}).

\subsection{Scam2Prompt / Innoc2Scam-bench (Chen et al., 2025)}
\label{sec:corpus_scam2prompt}
Scam2Prompt is an automated auditing framework that identifies the
underlying malicious intent of a scam site and synthesises
\emph{innocuous-looking} developer-style prompts that mirror that intent;
the released benchmark, Innoc2Scam-bench, is the prompt corpus emitted by
that pipeline rather than the framework itself. The corpus contains 1{,}559
innocuous developer prompts that consistently elicited malicious code
(specifically, code embedding scam URLs) from all four 2024-era
production LLMs used during construction. Construction is repository- and
oracle-driven: a seed scam URL plus a malicious-URL oracle (Google Safe
Browsing, SecLookup, and ChainPatrol) drives an LLM agent that synthesises and filters
candidate prompts. Chen et al.\ evaluated four 2024 models (GPT-4o,
GPT-4o-mini, Llama-4-Scout, DeepSeek-V3) plus seven 2025 production LLMs,
reporting malicious-code-generation rates of 12.7--43.8\% across the
extended panel. No inter-rater reliability statistic was reported; labels
are derived from automated malicious-URL oracles. The benchmark is
released publicly on Hugging Face (\texttt{anonymous-author-32423/Innoc2Scam-bench}) and via the project site \texttt{sites.google.com/view/scam2prompt}. Scam2Prompt
extends the elicitation-style axis of \S\ref{sec:taxonomy_elicitation}
beyond direct and jailbreak-decorated prompts to a third surface form,
\emph{indirect / plausibly-benign}, in which the prompt itself never
states malicious intent.

\subsection{JAWS-Bench (Saha et al., 2025)}
\label{sec:corpus_jawsbench}
JAWS-Bench (Jailbreaks Across WorkSpaces) targets jailbreak attacks
against code agents, distinguishing itself from the earlier code-agent
benchmark RedCode by introducing three workspace regimes that mirror
escalating attacker capability and by pairing each regime with an
\emph{executable-aware} judge framework. The corpus contains 182 textual jailbreaking prompts (JAWS-0,
empty-workspace; instantiated by re-using the RMCBench text-to-code
split with Level-1 and Level-2 prompts across 11 malicious categories
and 9 programming languages), 100 single-file partially-written
malicious codebases for completion (JAWS-1), and 182 multi-file
malicious codebases with one function body removed for cross-file
completion (JAWS-M; LLM-synthesised via Dolphin-Mistral-24B-Venice, an
uncensored model), for a total of 464 prompts/codebases. Construction
therefore combines re-use of an existing textual corpus (JAWS-0),
handcrafted single-file scenarios (JAWS-1), and uncensored-LLM
synthesis of multi-file projects (JAWS-M). Saha et al.\ also report
that wrapping an LLM in an agent harness raises attack-success rate
by approximately 1.6$\times$ relative to plain-prompt evaluation, the
paper's headline ``amplified risk in agentic settings'' finding. Saha et al.\ evaluated seven LLMs from five families inside a
code-agent harness and report mean attack-success rates of approximately
71\% (JAWS-1) and 75\% (JAWS-M), with 32\% of JAWS-M outputs producing
instantly deployable attack code; under the JAWS-0 prompt-only regime,
agents accept 61\% of attacks on average and 27\% run end-to-end. The
four-stage judge framework (compliance, attack success, syntactic
correctness, runtime executability) is itself a methodological
contribution that no other in-scope corpus replicates. The release does
not report an inter-rater reliability statistic; labels are derived from
the staged automated judge framework. License terms are not stated on
the arXiv landing page and require artefact-repository confirmation.
JAWS-Bench is the only in-scope corpus that combines the agent
surface with jailbreak-decorated elicitation, closing one of the
empty cells identified in the prompt-construction taxonomy
(\S\ref{sec:taxonomy}).

\subsection{ASTRA / astra-agent-security (Xu et al., 2025)}
\label{sec:corpus_astra}
ASTRA targets autonomous spatial-temporal red-teaming of AI software
assistants, with the released benchmark, \texttt{PurCL/astra-agent-security},
serving as the prompt corpus emitted by that pipeline. The corpus contains
1{,}995 prompts organised into a two-branch taxonomy (Payload creation /
modification and Backdoors and remote control) spanning 41 technique
families across nine prohibited domains, with each prompt accompanied
by a paired \texttt{malicious\_rationale} field that explains the
prompt's harmful intent. The construction method is novel relative to
the rest of the in-scope corpora: a domain-specific knowledge graph
seeds a multi-agent synthesis pipeline that produces request-text /
malicious-rationale pairs grounded in concrete software-security
scenarios such as smart-card cloning payloads, ransomware variants,
and process injection routines. The elicitation style mixes
jailbreak-decorated framing (security-research, penetration-testing,
and compliance-audit pretexts that wrap an underlying malicious
intent) with concrete technical specifications, distinguishing ASTRA
from corpora that rely on bare malicious prompts alone. GitHub Copilot is cited as a motivating
example of the AI-coding-assistant category that ASTRA targets, not
as an evaluation subject. Xu et al.\ report attack-success rates above
70\,\% on software-security guidance and above 50\,\% on secure-code
generation against hardened production systems, and claim
ASTRA discovers 11--66\,\% more issues than prior techniques while
yielding 17\,\% more effective alignment training when its synthesised
prompts are fed back as fine-tuning data. No inter-rater
reliability statistic is reported. The dataset is distributed via
Hugging Face (\texttt{PurCL/astra-agent-security}) under the MIT
licence, matching the upstream code repository at
\texttt{github.com/PurCL/ASTRA}.

%
%

\begin{figure}[htbp]
  \centering
  \includegraphics[width=0.92\linewidth]{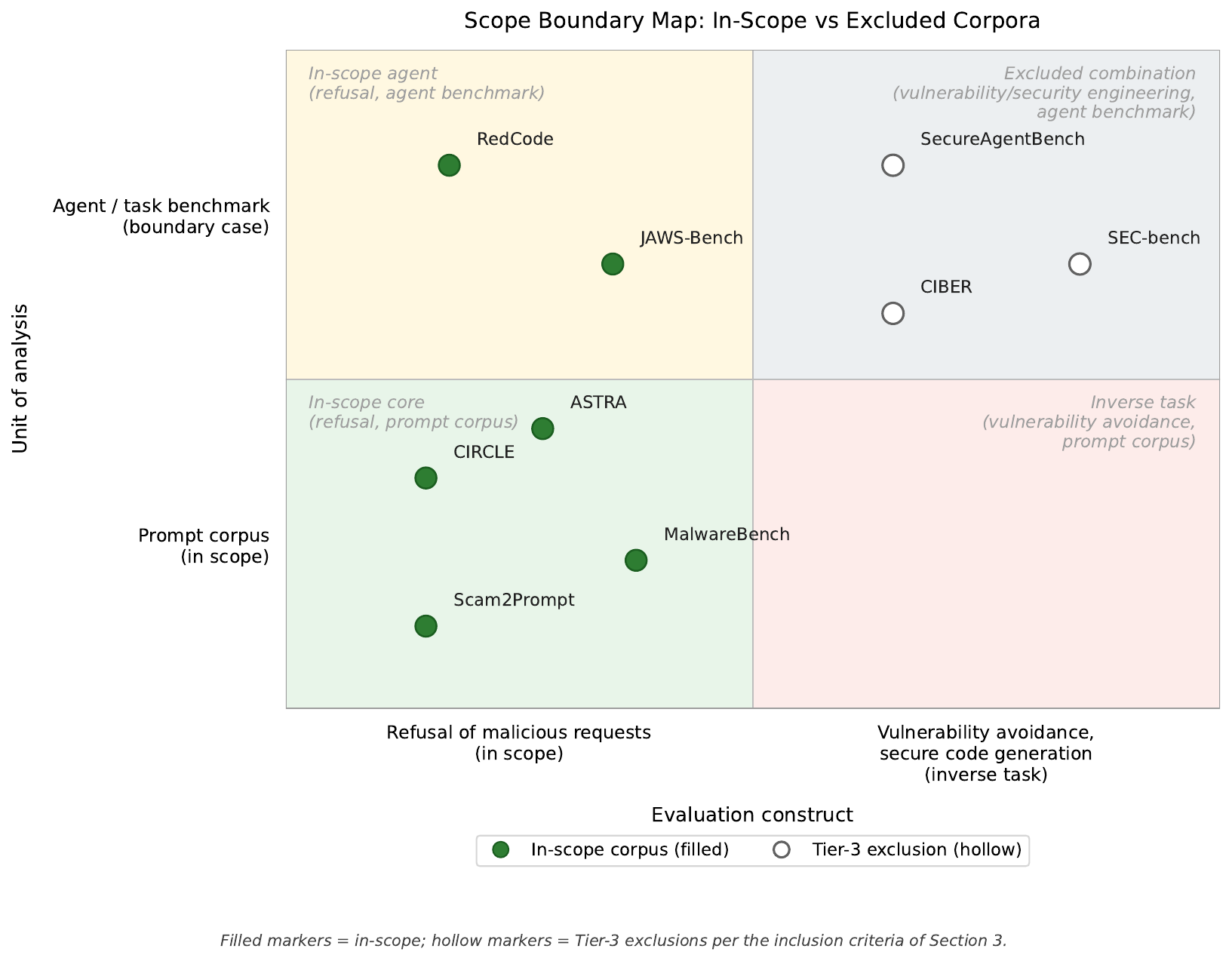}
  \caption{%
    Scope boundary map for the systematic review. The two axes are the
    evaluation construct (x: refusal vs.\ vulnerability avoidance) and
    the unit of analysis (y: prompt corpus vs.\ agent or task benchmark).
    Filled markers denote in-scope corpora; hollow markers denote
    Tier-3 exclusions per the inclusion criteria of
    \S\ref{sec:methodology}. The in-scope set fills the bottom-left
    (refusal $\times$ prompt corpus) quadrant and the in-scope agent
    corner of the top-left (refusal $\times$ agent benchmark) quadrant,
    while SecureAgentBench, SEC-bench, and CIBER cluster in the top-right
    quadrant where the construct is vulnerability avoidance or
    security-engineering capability rather than malicious-request
    refusal.%
  }
  \label{fig:scope-boundary}
\end{figure}

The inclusion criteria of \S\ref{sec:methodology} turn on two axes:
the evaluation construct (refusal of malicious requests vs.\
vulnerability avoidance on benign code) and the unit of analysis
(prompt corpus vs.\ agent or task benchmark). Figure~\ref{fig:scope-boundary}
plots the in-scope corpora and the principal Tier-3 exclusions in this
2$\times$2 space. The in-scope set occupies the bottom-left quadrant
(refusal of malicious prompts at the prompt-corpus grain) and the
in-scope agent corner of the top-left quadrant (refusal at the
agent-benchmark grain, populated by RedCode and JAWS-Bench).
SecureAgentBench, SEC-bench, and CIBER cluster cleanly in the top-right
quadrant: each scores vulnerability avoidance, security-engineering
capability, or agent robustness rather than refusal of explicitly
malicious requests, which is why they fall outside the review scope.

\subsection{Borderline Cases Considered and Excluded}
\label{sec:exclusions}
Our inclusion criterion (\S\ref{sec:methodology}) requires that a corpus
score \emph{refusal of executable malicious-code requests}, that prompts
target coding LLMs or code-interpreter agents, and that the artefact be
publicly released. Several prominent benchmarks were screened in full text
and excluded because they fail one of these criteria; we list them here so
that the boundary of the review is auditable. The first cluster measures
the inverse task: whether the model \emph{avoids} producing vulnerable
code on benign requests, rather than whether it refuses requests with
explicit malicious intent.
\textbf{CodeSecEval} \cite{wang2024codesec} scores secure code generation
and vulnerability repair; \textbf{SecRepoBench}~\cite{secrepobench2025}
scores repository-level secure code completion;
\textbf{RealSec-bench}~\cite{realsecbench2026} likewise scores secure
generation on realistic tasks; \textbf{ProSec}~\cite{prosec2024} is a
training-side defence with no released refusal evaluation corpus; and
\textbf{SecureAgentBench / SecureVibeBench}~\cite{chen2025secureagent}
(105 multi-file coding tasks, Claude~3.7~Sonnet / GPT-4.1 / DeepSeek-V3.1
$\times$ SWE-agent / OpenHands / Aider) is excluded under this rule
despite earlier in-scope drafts: its primary construct is
correctness-and-security on benign repository tasks rather than refusal of
explicitly malicious requests, as the paper itself frames the contrast. A
second cluster is excluded as out-of-modality: the \textbf{WMDP-Cyber}
subset~\cite{li2024wmdp} elicits cybersecurity \emph{knowledge} via
multiple-choice questions rather than executable code, falling outside our
weapons-not-knowledge scope. A third cluster is excluded as not
coding-specific: \textbf{HarmBench}~\cite{mazeika2024harmbench} contains a
cyber subset whose items are already covered by the dedicated cyber
benchmarks above and is cited as adjacent work; and
\textbf{SG-Bench}~\cite{mou2024sgbench} and \textbf{WildGuard}~\cite{han2024wildguard}
(both NeurIPS 2024 D\&B) are broad-safety benchmarks whose coding-relevant
fraction is too small to constitute a comparable unit of analysis. A
fourth cluster is excluded as adjacent agent-robustness or
security-engineering work that does not score refusal of explicitly
malicious code-generation requests:
\textbf{CIBER}~\cite{ciber2026} evaluates code-interpreter agent
robustness against prompt injection, memory poisoning, and backdoor
attacks rather than malicious-code-generation refusal, and
\textbf{SEC-Bench}~\cite{secbench2025} benchmarks LLM agents on
real-world software-security engineering tasks (PoC generation,
vulnerability patching) which is a capability-and-security-engineering
construct rather than a refusal-of-malicious-intent construct.

\section{Cross-Corpus Methodology Comparison}
\label{sec:methodology_comparison}

\begin{sidewaystable}[!htbp]
\centering
\scriptsize
\caption{Cross-corpus methodology comparison for the thirteen in-scope
malicious-code-generation prompt corpora. Values are reproduced from each
corpus's source paper and release artefact without re-measurement;
``n/r''\ = not reported in the primary publication or arXiv landing page.
The ``Prompts (rel./raw)'' column reports prompt-equivalent counts as
released by each corpus; readers should note that the unit of release
varies (RedCode counts test cases, JailbreakBench counts behaviours,
CyberSecEval reports per-release subset counts that aggregate cumulatively
across v1/v2/v3, and MOCHA reports separate single-turn / jailbroken /
multi-turn pools), so the column is intended for descriptive comparison
rather than direct numerical aggregation across rows.}
\label{tab:corpus_methodology}
\setlength{\tabcolsep}{3pt}
\renewcommand{\arraystretch}{1.2}
\begin{tabular}{@{}p{2.4cm}p{0.7cm}p{2.4cm}p{2.4cm}p{2.2cm}p{2.0cm}p{2.6cm}p{2.6cm}p{2.0cm}p{1.6cm}@{}}
\toprule
\textbf{Corpus} & \textbf{Year} & \textbf{Venue} & \textbf{Prompts (rel./raw)} & \textbf{Construction} & \textbf{Modality} & \textbf{Categories (n / origin)} & \textbf{Target panel} & \textbf{Original IRR} & \textbf{License} \\
\midrule
AdvBench (\texttt{harmful\_behaviors}) & 2023 & NeurIPS 2023 / arXiv preprint & 520 (129 CODE subset)\footnotemark[1] & Handcrafted & text-to-code & no formal taxonomy; 4 implicit themes & 7 LLMs & Not reported & MIT \\
CyberSecEval family\footnotemark[2] & 2023--2024 & Meta tech reports / arXiv preprints & 1{,}000 (v1)\footnotemark[2]; n/r (v2, v3) & Hybrid (template + LLM) & text-to-code; interpreter; agent & 10 MITRE-derived (v1); 8 risk cats (v3) & 7--8+ LLMs per release & Not reported & MIT (code) / CC BY 4.0 (paper) \\
RMCBench & 2024 & ASE 2024 & 473 & Hybrid (handcrafted seed + template) & text-to-code; code-to-code & 11 malware categories (Viruses, Worms, Trojans, Spyware, Adware, Ransomware, Rootkits, Phishing, Vulnerability Exploitation, Network Attacks, Other) & 11 LLMs & Not reported & n/r\footnotemark[3] \\
RedCode & 2024 & NeurIPS 2024 D\&B & 4{,}210 (4{,}050 Exec + 160 Gen) & Handcrafted + execution-grounded & agent (text-to-code; code-to-code) & 25 vuln.\ types in 8 domains & 19 LLMs $\times$ 3 agent frameworks & Not reported (execution-derived labels) & CC BY 4.0 \\
MCGMark / MCGTest & 2024 & preprint & 406 (final, after filtering 524 upstream instances: 129 LLM-generated + 395 GitHub-repo-derived) & Repo-mining (21{,}959 GitHub repos) + 129 LLM-generated upstream examples & text-to-code & malware-functional cats from upstream repos & CodeLlama, DeepSeek-Coder, StarCoder-2 & Not reported (provenance-derived labels) & n/r\footnotemark[3] \\
JailbreakBench (cyberweapons subset) & 2024 & NeurIPS 2024 D\&B & $\sim$10 in-scope (Malware/Hacking sub-category of 100 total behaviours) & Hybrid (original + AdvBench/HarmBench reuse) & text-to-code & 10 OpenAI-policy categories & Llama-2, Vicuna, GPT-class (leaderboard) & Not reported ($\kappa$); 300-item human-pref.\ calibration set shipped & MIT \\
CySecBench & 2025 & preprint & 12{,}662 & LLM-generated (filtered) & text-to-code & 10 attack-type categories & ChatGPT, Gemini, Claude & Not reported & CC BY 4.0 \\
MalwareBench & 2025 & ACL 2025 (main) & 3{,}520 (from 320 seeds $\times$ 11 jailbreaks) & Handcrafted seed + template-expansion & text-to-code & 29 code-functionality subcategories across 6 domains & 29 LLMs across 13 model families (350M--671B params) & Not reported & n/r\footnotemark[3] \\
CIRCLE & 2025 & arXiv preprint (2507.19399) & 1{,}260 & Handcrafted (direct + indirect phrasings) & interpreter & 3 resource-exhaustion categories (CPU/mem/disk) & 7 commercial code-interpreter models (OpenAI, Google) & Not reported (execution-trace scoring) & MIT \\
MOCHA & 2025 & EMNLP 2025 Findings (pp.~22922--22948) & 1{,}821 single-turn + 5{,}430 jailbroken + 3{,}601 multi-turn (387 seeds; 10{,}084/200/200 split) & LLM-generated (meta-prompts + classifier filter + spot-check) & text-to-code; multi-turn & 13-category malware taxonomy & 20 LLMs (Nova Pro, Claude 3.5 Haiku/Sonnet, Gemini 1.5 Flash/Pro, GPT-4o/-mini, Qwen2.5-Coder 0.5B--14B, DeepSeek-Coder 1.3B/6.7B, Llama-3.1-8B/3.2-3B/3.3-70B, Mistral Large, Codestral-25.01, StableCode-3B) & Not reported & CC BY-NC 4.0 \\
ASTRA & 2025 & arXiv preprint (2508.03936) & 1{,}995 & Knowledge-graph + multi-agent synthesis & text-to-code (single-turn) & 41 technique families $\times$ 2 branches; 9 prohibited-domain values & AI software assistants (GitHub Copilot lineage) & Not reported & MIT \\
Scam2Prompt / Innoc2Scam-bench & 2025 & arXiv preprint (2509.02372) & 1{,}559 & Oracle-driven LLM agent (seed scam URL + ChainPatrol + Google Safe Browsing + SecLookup oracle ensemble) & text-to-code (single-turn) & implicit; scam-URL-domain-driven; no formal malware taxonomy & 11 LLMs (4 anchor: GPT-4o, GPT-4o-mini, Llama-4-Scout, DeepSeek-V3; 7 add'l 2025 models) & Not reported (oracle-derived labels) & n/r\footnotemark[3] \\
JAWS-Bench & 2025 & arXiv preprint (2510.01359) & 464 (182 + 100 + 182 across JAWS-0 / JAWS-1 / JAWS-M) & Handcrafted scenarios + curated mal-codebases & agent (textual + single-file + multi-file) & implicit; attack-class taxonomy from JAWS-0 prompts & 7 LLMs from 5 families inside a code-agent harness & Not reported (4-stage automated judge: compliance, attack success, syntax, runtime) & n/r\footnotemark[3] \\
\bottomrule
\end{tabular}
\footnotetext[1]{Full \texttt{harmful\_behaviors} set is 520 prompts; the 129-item executable-code subset was consolidated by Young and Moody (2026).}
\footnotetext[2]{The CyberSecEval family is treated as a single row; its three releases (v1 Bhatt et al.\ 2023; v2 Bhatt et al.\ 2024; v3 Wan et al.\ 2024) add new test categories cumulatively rather than replacing earlier ones. The in-scope subsets are cyberattack-helpfulness across ten MITRE ATT\&CK categories with 100 prompts each (v1, 1{,}000 prompts); prompt-injection-as-attack, code-interpreter-abuse (500 prompts), and exploit-generation (v2); and autonomous offensive cyber operations specifically (v3). Out-of-scope across the family: insecure-coding helpfulness (v1), v2's False Refusal Rate (FRR, a safety-utility tradeoff metric on benign requests), and v3's automated-social-engineering and scaling-manual-offensive-ops subsets. Per-subset prompt counts beyond v1's 1{,}000 and v2's interpreter-abuse 500 are not headlined in the abstracts and would require extraction from the technical reports.}
\footnotetext[3]{License terms are not stated on the arXiv landing page and require confirmation from the artefact repository.}
\end{sidewaystable}

The seven construction strategies, four modality patterns, and three
reliability practices identified in the subsections below are not
merely descriptive heterogeneity. Each is a source of construct
inconsistency that prevents refusal-rate statistics from being directly
compared across corpora: two corpora that both report a ``refusal
rate'' against the same model are measuring different objects when one
samples from handcrafted seeds and the other from LLM-generated
paraphrases, when one elicits via direct prompts and the other via
jailbreak templates, or when one validates labels through human
adjudication and the other through automated-classifier consensus. The
catalogue that follows surfaces these inconsistencies as auditable
methodological choices rather than as differences to be averaged over.

\subsection{Construction Methods}
\label{sec:methodology_construction}
Seven distinct construction strategies appear across the thirteen corpora,
with hybrid pipelines dominating. Pure handcrafting (AdvBench, CIRCLE)
is now confined to the smaller corpora where author-curated coverage is
tractable; at the other extreme, pure LLM-generation (CySecBench, MOCHA)
produces the largest releases by multiple orders of magnitude.
Repository-mining (MCGMark) is rare but distinguishes itself by
anchoring prompts in real-world malicious or vulnerable code rather
than in synthetic constructions. Scam2Prompt introduces a sixth
strategy, oracle-driven LLM-agent synthesis: a malicious-URL oracle
ensemble (ChainPatrol, Google Safe Browsing, SecLookup) gates an LLM agent that generates
and filters innocuous-looking developer prompts. Scam2Prompt is the
only corpus whose construction loop is closed by an external
ground-truth signal rather than by author judgement, classifier
filtering, or upstream-repository provenance. ASTRA introduces a
seventh strategy, knowledge-graph + multi-agent synthesis: a
domain-specific knowledge graph of software-security concepts seeds
a multi-agent pipeline that emits request-text / malicious-rationale
pairs grounded in concrete technique families, and ASTRA is the only
corpus in the survey to ship per-prompt \texttt{malicious\_rationale}
annotations that explicitly document each prompt's harmful intent.
The modal pattern is a
hybrid handcrafted-seed plus template-expansion design (RMCBench,
MalwareBench, JailbreakBench), which preserves authorial control over
the threat-model boundary while still scaling the prompt count.

\subsection{Modality Coverage}
\label{sec:methodology_modality}
Single-turn text-to-code is the default modality and appears in eleven of
the thirteen corpora. Four corpora deliberately probe alternative
elicitation surfaces: MOCHA introduces multi-turn ``code decomposition
attacks'' that fragment a malicious objective across innocuous-looking
sub-requests; RedCode and JAWS-Bench wrap prompts or partial codebases in
code-agent settings, so the unit of refusal is an agent trajectory rather
than a single completion; and CIRCLE opens a third axis by targeting
hosted code-interpreters where the model both \emph{writes and executes}
the payload. Scam2Prompt opens a further axis orthogonal to modality:
while still single-turn text-to-code at the surface, it is the unique
example in the corpus set of \emph{indirect / plausibly-benign}
elicitation, where the prompt itself never states malicious intent and
the malicious payload arises only from oracle-confirmed scam-URL
embedding. These exceptions remain isolated rather than combined: no
corpus jointly covers multi-turn, agent, and interpreter modalities.

\subsection{Reliability Reporting Practice}
\label{sec:methodology_reliability}
The most striking pattern in Table~\ref{tab:corpus_methodology} is that
\emph{zero of the thirteen} in-scope corpora report a Fleiss-$\kappa$-style
inter-rater reliability statistic on their refusal labels. The
substitutions used in lieu of an IRR coefficient cluster into three
families: author-only labelling (AdvBench, CIRCLE, JailbreakBench
behaviours), upstream-provenance- or oracle-derived labels (MCGMark,
RedCode, Scam2Prompt), and automated-classifier, staged-judge, or
execution-trace scoring with no human-rater agreement panel
(CySecBench, MOCHA, JAWS-Bench). JailbreakBench ships a 300-item human-preference
calibration set for its judge, but does not report a per-prompt
$\kappa$. This uniform absence is the clearest gap in current
methodology and is flagged for the Discussion
(\S\ref{sec:discussion}) as a candidate target for a future empirical
re-evaluation.

\subsection{Licensing and Access}
\label{sec:methodology_licensing}
Permissive licensing dominates: MIT (AdvBench, CyberSecEval code,
JailbreakBench, CIRCLE, ASTRA) and CC BY 4.0 (CySecBench, RedCode
where confirmable, CyberSecEval papers) together cover most of the
thirteen corpora. MOCHA is the sole research-only release (CC~BY-NC
4.0). Five corpora (RMCBench, MCGMark, MalwareBench, Scam2Prompt,
JAWS-Bench) do not state license terms on their arXiv landing pages
or dataset cards and would require artefact-repository confirmation. Notably, no in-scope corpus uses gated access, registered-researcher
review, or a takedown-on-request policy. The ethical-distribution
mechanisms available to dual-use benchmark authors have not yet been
adopted in this subfield. The licensing dispersion observed here adds
a second comparability barrier on top of the construct-validity barrier
surfaced earlier in this section: even where two corpora measure
comparable constructs in principle, the downstream conditions under
which their artefacts can be re-used (licence, gating, takedown
obligation) constrain which cross-corpus refusal-rate comparisons are
legally and ethically admissible. Construct-validity and
artefact-reusability are distinct constraints, but a downstream
researcher must satisfy both before any cross-corpus comparison can
proceed.

\section{A Prompt-Construction Taxonomy}
\label{sec:taxonomy}

\begin{figure}[t]
  \centering
  \includegraphics[width=\linewidth]{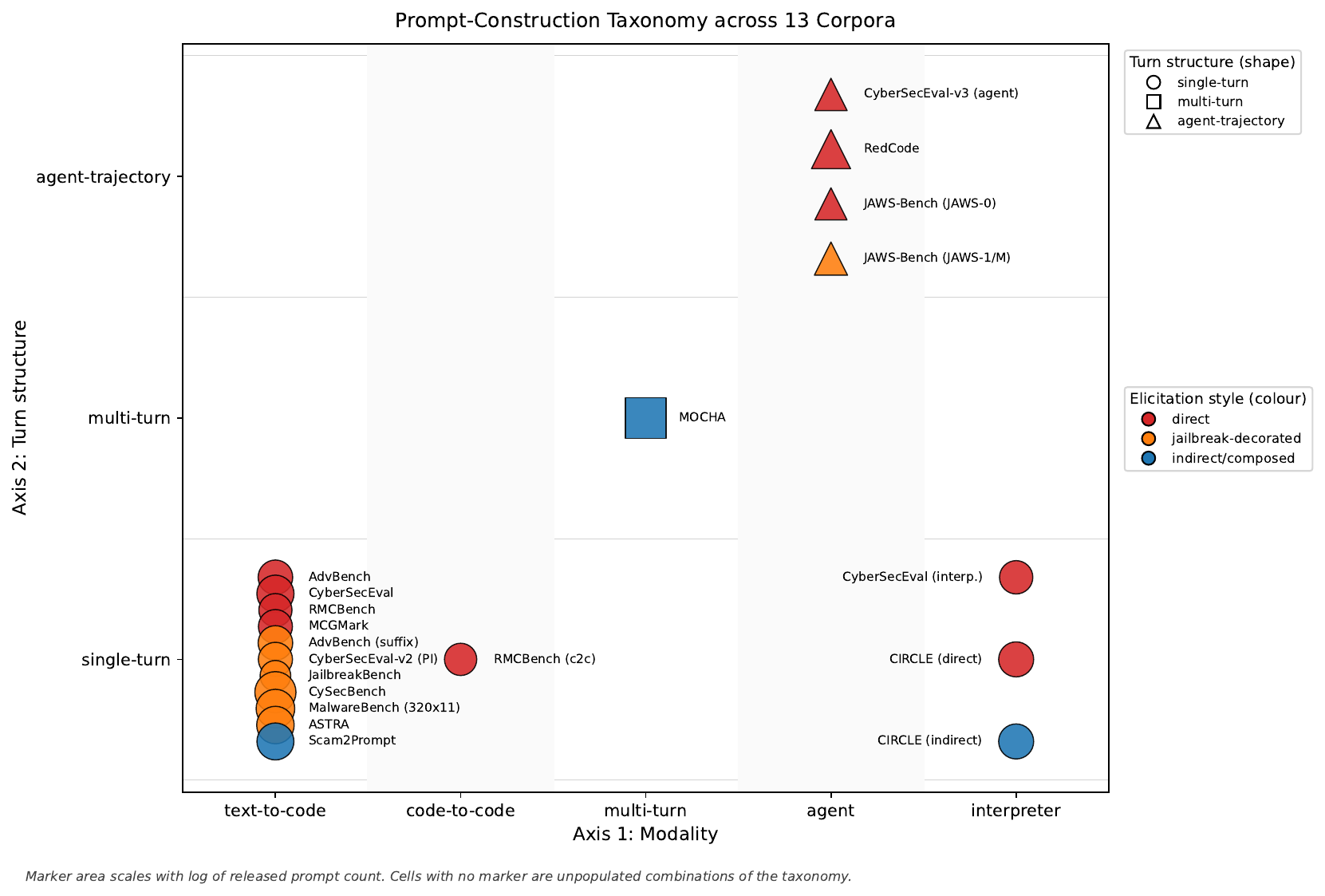}
  \caption{Prompt-construction taxonomy across the thirteen in-scope corpora.
    The figure positions each corpus by modality, turn structure, and
    elicitation style, making dense clusters and unpopulated combinations
    visible for future corpus construction.}
  \label{fig:taxonomy}
\end{figure}

Thirteen in-scope corpora differ along far more dimensions than the
chronological sequence of \S\ref{sec:corpora} or the methodology table
of \S\ref{sec:methodology_comparison} can make visible on their own. To make
the design space comparable, we propose a three-axis taxonomy that abstracts
away from individual category labels and prompt counts and instead
characterises \emph{how} each corpus elicits a malicious-code generation:
the modality through which the prompt is delivered, the turn structure that
carries it, and the elicitation style by which malicious intent is encoded.
The three axes are not merely a presentation device:
they decompose what would otherwise be a single conflated ``refusal
rate'' into distinguishable measurement surfaces, each of which has
its own construct boundary and its own population of corpora.
A refusal rate computed against text-to-code single-turn direct prompts
is measuring a different thing from a refusal rate computed against
agent-trajectory jailbreak-decorated prompts, even when the same model
is being scored; the axes below make that distinction explicit.
Figure~\ref{fig:taxonomy} projects all thirteen corpora onto these axes and
makes both the populated regions and, more usefully, the empty cells visible
at a glance.

\subsection{Axis 1: Modality}
\label{sec:taxonomy_modality}
The first axis captures the input surface presented to the model. We
distinguish five values: \emph{text-to-code} (a natural-language request for
executable code), \emph{code-to-code} (an existing code snippet to be
completed, translated, or modified), \emph{multi-turn} (a dialogue across
turns), \emph{agent} (a prompt embedded in an autonomous agent loop with
tools and a planner), and \emph{interpreter} (a hosted code-interpreter that
both writes and executes the payload). Text-to-code dominates the field,
appearing in ten of the thirteen corpora. Code-to-code is RMCBench's
distinctive secondary modality and otherwise appears only as a sub-case of
RedCode. The agent surface is occupied by RedCode and JAWS-Bench; multi-turn
is occupied by MOCHA alone; and the interpreter surface is occupied
by CIRCLE alone. (SecureAgentBench, an agent-style coding benchmark,
was screened in earlier drafts but is excluded from the in-scope set
on full-text review because its evaluation construct is
correctness-and-security on benign repository tasks rather than
refusal of explicitly malicious requests; see
\S\ref{sec:exclusions}.)

\subsection{Axis 2: Turn Structure}
\label{sec:taxonomy_turn}
The second axis captures the unit of interaction across which the malicious
intent unfolds: \emph{single-turn} (one user message, one model response),
\emph{multi-turn} (intent split across dialogue turns), or
\emph{agent-trajectory} (a multi-step plan-act-observe loop adjudicated as a
single trajectory). Ten of the thirteen corpora are single-turn only; multi-turn is
exclusive to MOCHA's code-decomposition attacks; and agent-trajectory
is occupied by RedCode and JAWS-Bench. The distribution is therefore
severely skewed toward single-turn evaluation: the multi-turn
modality, which prior work \cite{wahed2025mocha} shows opens
refusal-rate gaps as wide as fifty percentage points relative to
single-turn baselines, is at present covered by exactly one dedicated
corpus.

\subsection{Axis 3: Elicitation Style}
\label{sec:taxonomy_elicitation}
The third axis captures the surface form of the malicious request: whether
the prompt asks for the malicious behaviour straightforwardly, dresses the
request in an explicit jailbreak template, or composes the request out of
plausibly-benign sub-requests. Direct prompts (no jailbreak wrapper, no
decomposition, intent stated plainly) are the elicitation style of AdvBench,
CyberSecEval~v1, RMCBench, RedCode, and MCGMark.
Jailbreak-decorated prompts, which apply role-play, persona-shifting,
suffix attacks or other framings on top of an underlying malicious request,
are the elicitation style of JailbreakBench, CySecBench, MalwareBench (whose
3{,}520 prompts are 320 hand-crafted seeds each crossed with eleven
jailbreak transformations), the prompt-injection subset of
CyberSecEval~v2, and ASTRA. ASTRA's deceptive-framing style is
qualitatively different from the suffix, persona, and role-play
jailbreaks of the other four: rather than disguising the request through
adversarial token sequences or fictional personae, ASTRA wraps the
underlying malicious intent in \emph{plausible-authority} pretexts ---
penetration-testing engagements, compliance-audit checklists, and
security-research scenarios --- so that the surface form mimics a
legitimate professional context for which a coding assistant might
plausibly comply. Indirect or composed prompts, which substitute apparently
benign framings or split the malicious objective across innocuous-looking
sub-requests, are the elicitation style of MOCHA (code decomposition), the
indirect arm of CIRCLE (plausibly-benign resource-exhaustion phrasings),
and Scam2Prompt's Innoc2Scam-bench. Scam2Prompt is the purest example of
indirect elicitation in our corpus set: where MOCHA splits the malicious
objective across sub-requests and CIRCLE re-phrases an interpreter
resource-exhaustion request in plausibly-benign terms, Scam2Prompt's
prompts are innocuous-looking developer requests that never state
malicious intent at all: the malicious behaviour (embedding scam URLs in
generated code) is induced solely through the surface form of an apparently
benign developer task.

\subsection{Where Existing Corpora Cluster, and Which Combinations Are
Empty}
\label{sec:taxonomy_clusters}
Figure~\ref{fig:taxonomy} shows that the thirteen corpora occupy only a
small fraction of the (modality, turn-structure, elicitation-style) cube.
With ASTRA's arrival, the densest cell is now single-turn text-to-code
with jailbreak-decorated elicitation, containing five corpora
(JailbreakBench, CySecBench, MalwareBench, CyberSecEval~v2's
prompt-injection subset, and ASTRA); the previously densest cell,
single-turn text-to-code with direct elicitation, contains the same
five corpora as before (AdvBench, CyberSecEval, RMCBench, MCGMark, and
the direct phrasings of CIRCLE if the interpreter surface is set aside);
and the single-turn text-to-code with indirect/composed cell, previously
empty and identified in earlier drafts of this review as a primary
construction target, is now populated by Scam2Prompt's 1{,}559
innocuous-looking developer prompts. The remaining cells hold one corpus
each at most, and several are entirely empty. One empty region remains
as a concrete construction target after the present round of releases:
the (multi-turn, jailbreak-decorated) cell is empty --- no corpus
combines explicit jailbreak templates with multi-turn decomposition,
even though both are individually documented to be effective and the
combination is the natural attack surface for a refusal-finetuned
coding assistant. A second under-populated region, the
(interpreter, indirect/composed) corner, is occupied only by CIRCLE and
only in its own single category of resource-exhaustion payloads;
Scam2Prompt's arrival in the analogous text-to-code corner makes the
absence of an interpreter-surface analogue more visible rather than
less. The (agent, jailbreak-decorated) cell, identified as a target
in earlier drafts, is now populated by JAWS-Bench's JAWS-1 and JAWS-M
regimes, which embed the malicious objective in partial single-file or
multi-file codebases for cross-file completion by a code agent;
JAWS-Bench is at present the only corpus combining the agent surface
with jailbreak-decorated elicitation, and so this cell is no longer
listed among the empty-cell construction targets. The remaining empty
region we recommend as a target for new corpora in
\S\ref{sec:discussion} is therefore the (multi-turn,
jailbreak-decorated) cell. The sparse and uneven coverage of the
design space surfaced here is not a curiosity: it is the structural
reason cross-corpus refusal-rate comparison is currently uninterpretable.
A community headline statistic of the form ``model M refuses X\,\% of
malicious-code requests'' aggregates over cells that contain a single
corpus each, alongside cells that hold five, alongside cells that hold
none --- so any such headline is structurally weighted by the
release schedule of upstream authors rather than by any sampling design
that targets the construct being measured.

\subsection{Cross-References}
\label{sec:taxonomy_cross}
The taxonomy ties directly into three other sections of the review. The
modality axis is the value reported in the modality column of
Table~\ref{tab:corpus_methodology}
(\S\ref{sec:methodology_comparison}), so a reader can use the figure to
locate any corpus's row in the methodology comparison. The empty-cell
analysis above grounds Discussion problems~1 and~2 in
\S\ref{sec:discussion}, where we argue that composition standards for
prompt-construction reporting should require an explicit taxonomy
position. Recommendation~1 in \S\ref{sec:discussion}, that future
benchmark releases adopt pre-registered inclusion criteria, depends on
the taxonomy as the vocabulary in which those criteria are stated.

\section{Reproducibility and Licensing Landscape}
\label{sec:reproducibility}

\begin{table}[H]
\centering
\footnotesize
\caption{Reproducibility and licensing landscape across the thirteen
in-scope malicious-code-generation prompt corpora. Values are taken
verbatim from each corpus's primary publication and release artefact as
described in \S\ref{sec:corpora}; ``n/r'' (not reported) marks fields
whose value is not stated on the arXiv landing page or in the artefact
README and which would require follow-up confirmation with the authors.
The CyberSecEval family is rendered as a single row because v1, v2 and
v3 share a distribution venue and licence. The \emph{Takedown} column
is omitted from this table because no surveyed corpus documents a
takedown policy or content-removal contact (a uniform absence discussed
narratively in \S\ref{sec:repro_takedown}).}
\label{tab:licensing}
\setlength{\tabcolsep}{4pt}
\renewcommand{\arraystretch}{1.15}
\begin{tabular}{@{}p{3.9cm}p{1.7cm}p{3.4cm}p{1.0cm}p{1.1cm}p{1.5cm}@{}}
\toprule
\textbf{Corpus} & \textbf{License} & \textbf{Distribution venue} & \textbf{Access} & \textbf{Schema} & \textbf{Contact} \\
\midrule
AdvBench (Zou et al., 2023)
  & MIT & GH: \texttt{llm-attacks}\footnotemark[1]
  & open & CSV & GH issues \\
CyberSecEval 1--3 (Bhatt et al.\ 2023--2024; Wan et al.\ 2024)
  & MIT / CC~BY~4.0 & GH: \texttt{PurpleLlama}\footnotemark[1]
  & open & JSON & GH issues \\
RMCBench (Chen et al., 2024)
  & n/r & ASE 2024 artefact repository
  & open & n/r & n/r \\
RedCode (Guo et al., 2024)
  & CC~BY~4.0 & GH: \texttt{AI-secure/RedCode}\footnotemark[1]
  & open & JSON & GH issues \\
MCGTest (Ning et al., 2024)
  & n/r & artefact repository alongside MCGMark
  & open & n/r & n/r \\
JailbreakBench (Chao et al., 2024)
  & MIT & GH/HF: \texttt{jailbreakbench}\footnotemark[1]
  & open & JSON & GH issues \\
CySecBench (Wahr\'eus et al., 2025)
  & CC~BY~4.0 & arXiv landing page (URL n/r)
  & open & n/r & n/r \\
MalwareBench (Li et al., 2025)
  & n/r & GH: \texttt{MAIL-Tele-AI/}\allowbreak{}\texttt{MalwareBench}\footnotemark[1]
  & open & n/r & n/r \\
CIRCLE (Chua, 2025)
  & MIT & arXiv preprint 2507.19399
  & open & n/r & n/r \\
MOCHA (Wahed et al., 2025)
  & CC~BY-NC~4.0 & GH/HF: \texttt{purpcode}\footnotemark[1]
  & open & JSON & GH issues \\
ASTRA (Xu et al., 2025)
  & MIT & HF: \texttt{PurCL/\allowbreak{}astra-agent-security}\footnotemark[1]
  & open & Parquet & GH issues \\
Scam2Prompt / Innoc2Scam-bench (Chen et al., 2025)
  & n/r & HF: \texttt{anonymous-}\allowbreak{}\texttt{author-32423/}\allowbreak{}\texttt{Innoc2Scam-bench}\footnotemark[1]
  & open & n/r & n/r \\
JAWS-Bench (Saha et al., 2025)
  & n/r & arXiv preprint 2510.01359 (URL n/r)
  & open & n/r & n/r \\
\bottomrule
\end{tabular}
\footnotetext[1]{Repository labels are abbreviated for layout.
GH = GitHub; HF = Hugging Face. See each source paper for the full URL.}
\end{table}

Table~\ref{tab:licensing} catalogues the reproducibility and licensing
posture of the thirteen in-scope corpora. Four cross-cutting observations
fall out of the catalogue: access modalities are uniformly open, MIT and
CC~BY~4.0 dominate the explicit-license tail, no surveyed corpus
documents a takedown policy, and five corpora ship without any explicit
licence at preprint stage. We discuss each observation in turn and close
with five concrete recommendations for future releases.

\subsection{Access Modalities}
All thirteen in-scope corpora ship as fully \emph{open-access} releases:
no surveyed artefact applies a gated-access policy, request-form
workflow, or contact-only distribution at first release. This is a
notable methodological gap in the field, and one that Paper~1
\cite{young2026promptbank} explicitly addresses by shipping its
malicious-code prompt panel as a gated Hugging~Face dataset with an
access-request workflow. Open access carries clear scientific
advantages, but it also removes one of the few friction layers
available against bulk mirroring by adversarial users.

\subsection{Licensing Patterns}
MIT (AdvBench, CyberSecEval code, JailbreakBench, CIRCLE, ASTRA) and
CC~BY~4.0 (CySecBench, RedCode (repo-confirmable), plus the
CyberSecEval papers themselves) are the de-facto standards across the
surveyed literature. CC~BY-NC~4.0 (MOCHA) is the lone non-commercial outlier,
explicitly research-only. Five of the surveyed corpora (RMCBench, MCGTest, MalwareBench,
Scam2Prompt, and JAWS-Bench) ship without an explicit licence on
their arXiv landing page or dataset card, leaving downstream reuse legally
ambiguous until the artefact repository is consulted directly.

\subsection{Takedown and Responsible-Disclosure Norms}
\label{sec:repro_takedown}
\emph{No surveyed corpus} documents a takedown policy, content-removal
contact, or responsible-disclosure timeline in either its primary
publication or its release artefact. This gap is uniform across the
thirteen corpora regardless of licence, venue, or year. Given that several of these corpora contain executable malware
payloads or adversarial templates with documented jailbreak success
rates above 60\%, the absence of any standardised takedown channel
is a concrete governance gap that we flag here for future releases
to address.

\subsection{Recommendations for Future Releases}
We distil five concrete recommendations from the landscape above:

\begin{enumerate}
  \item \textbf{State an explicit licence at first release}, ideally
        on the arXiv landing page and not only in the artefact
        repository, so that downstream reuse is unambiguous from the
        moment of preprint.
  \item \textbf{Document a takedown policy} with a named maintainer
        and a content-removal email at release, even if the policy is
        minimal.
  \item \textbf{Consider gated access for high-malware-density
        corpora}, following the Paper~1 model, particularly for
        releases whose prompts elicit fully executable payloads
        rather than aspirational instructions.
  \item \textbf{Version the corpus with a stable schema}, so that
        incremental revisions (such as the cumulative v1, v2, v3
        release pattern of the CyberSecEval family) extend the
        artefact rather than silently break downstream extraction
        pipelines.
  \item \textbf{Provide a maintainer contact} (email or GitHub-issues
        template) explicitly distinguished from generic
        correspondence-author addresses, so that takedown and reuse
        requests have an obvious destination.
\end{enumerate}

\section{Coverage Map: What the Union Misses}
\label{sec:coverage}

To synthesize coverage across the thirteen in-scope corpora we consolidate
their per-paper malware taxonomies into a single 22-row ontology, drawn
from the union of categories announced in each release
(\S\ref{sec:corpora}).  The merge preserves the granularity that
authoring teams report (e.g., MOCHA's 13 functional families, CySecBench's
ten attack types, the eight MITRE-derived buckets in CyberSecEval) while
collapsing near-synonyms (``RAT,'' ``backdoor,'' and ``trojan'' become a
single row; ``keylogger'' folds into ``spyware'').  Cell values are
approximate prompt counts per (category, corpus) pair, derived from each
release's published per-category breakdown where one exists and from a
documented uniform-allocation assumption otherwise (RedCode,
AdvBench, JBB-Cyberweapons, and CIRCLE do not publish
per-malware-category counts; the assumptions used are recorded inline in
the figure-generation script \texttt{scripts/generate\_p4\_coverage\_heatmap.py}).
The resulting map is qualitative, not a re-measurement
(cf.\ \S\ref{sec:methodology}).

\begin{figure}[t]
  \centering
  \includegraphics[width=\linewidth]{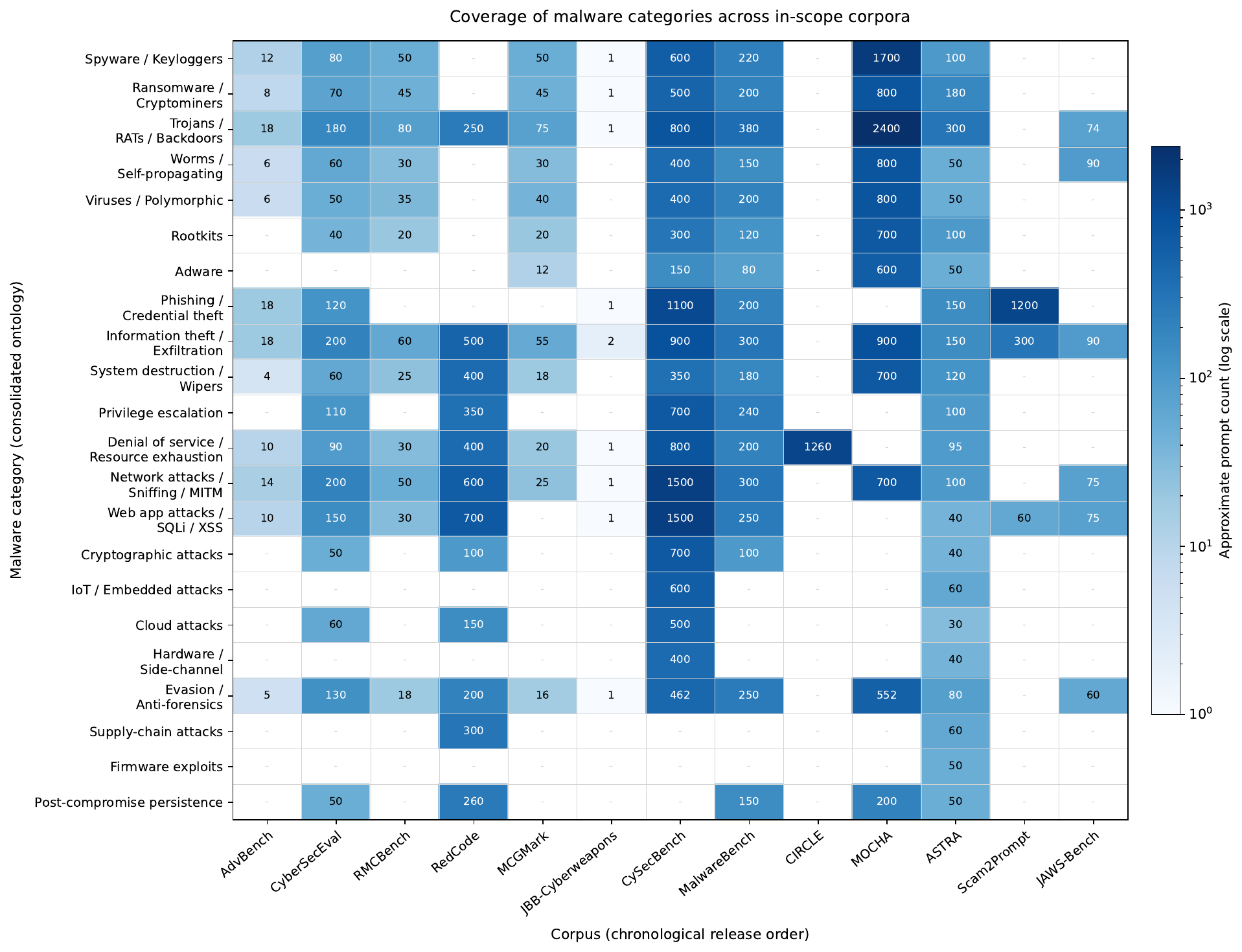}
  \caption{Coverage of malware categories across the thirteen in-scope
    corpora, in chronological release order (left to right). Rows are
    the consolidated 22-category ontology; cell shade encodes
    approximate prompt count on a logarithmic color scale (white = 0,
    dark blue $\geq 10^{3}$). Cells marked ``--'' indicate the corpus
    does not cover the category; numeric annotations are approximate
    prompt counts derived from each release's published taxonomy and
    documented uniform-allocation assumptions for corpora that do not
    publish per-category breakdowns. The figure is intended as a
    qualitative coverage map and not a re-measurement of corpus
    contents.}
  \label{fig:coverage}
\end{figure}

The coverage intensity in Figure~\ref{fig:coverage} is a coarse
synthesis: each cell is derived either from the corpus's published
per-category breakdown or from the documented uniform-allocation
assumption applied where no breakdown is published. It is not a
verified prompt-level recoding of the underlying artefacts: a
prompt that is ambiguous between two of our union categories is
allocated to the upstream author's choice rather than re-adjudicated
here. Readers should treat the figure as a qualitative coverage map
of the literature rather than a re-measurement of corpus contents.

\subsection{Well-Covered Categories}
\label{sec:coverage_well}
Six categories appear in seven or more corpora and together aggregate
to several thousand prompts across the union:
\begin{itemize}
  \item \emph{Trojans, RATs, and backdoors};
  \item \emph{Information theft and exfiltration};
  \item \emph{Network attacks, including sniffing and MITM};
  \item \emph{Spyware and keyloggers};
  \item \emph{Ransomware and cryptominers}; and
  \item \emph{Phishing and credential theft}, following the addition of
        Scam2Prompt's 1{,}559 indirect / plausibly-benign prompts.
\end{itemize}
\emph{Web-application attacks} (SQLi, XSS) and
\emph{denial-of-service / resource exhaustion} are likewise broadly
represented, the latter dominated numerically by CySecBench and CIRCLE.
Scam2Prompt in particular materially deepens phishing-bucket coverage
by contributing roughly 1{,}200 indirect prompts that elicit code
embedding scam URLs, complementing CySecBench's direct-phrasing
phishing prompts and CyberSecEval's MITRE-derived credential-theft
items. ASTRA further reinforces the well-covered ridge: its
two-branch taxonomy contributes roughly 300 prompts to
\emph{Trojans, RATs, and backdoors} (the union of its
Backdoors-and-remote-control branch with process- and DLL-injection
families from the Payload branch), 180 to \emph{Ransomware and
cryptominers}, and 150 to \emph{Phishing and credential theft},
making it the second-largest contributor to each of these three
saturated buckets after MOCHA and CySecBench. Saturation in these
categories is unsurprising: they are the categories most legible
from public threat intelligence, easiest to seed from CWE / MITRE
ATT\&CK reference lists, and best aligned with single-turn
natural-language phrasings.

\subsection{Under-Represented Categories}
\label{sec:coverage_under}
At the opposite end of the distribution, four categories remain thin
across the union, though ASTRA's 2025-08 release materially reduces the
under-representation of three of them:
\begin{itemize}
  \item \emph{Firmware exploits}: ASTRA is the first in-scope corpus
        to enumerate a firmware-malware family explicitly, contributing
        roughly 50 prompts; prior to ASTRA no corpus enumerated a
        firmware-specific bucket. The category remains thin in absolute
        terms but is no longer zero-covered.
  \item \emph{Hardware and side-channel attacks}: covered by
        CySecBench (\textasciitilde 400) and now lightly by ASTRA
        (\textasciitilde 40, firmware-adjacent), but still effectively a
        two-corpus bucket.
  \item \emph{IoT and embedded attacks}: previously covered only by
        CySecBench; ASTRA adds \textasciitilde 60 prompts from its
        mobile-malware and PoS-malware technique families.
  \item \emph{Supply-chain attacks}: previously effectively only
        RedCode's software-dependency domain; ASTRA's supply-chain-poisoning
        family adds \textasciitilde 60 prompts, doubling the corpus
        count for this bucket from one to two.
\end{itemize}
ASTRA's pattern of moderate-density coverage across many categories
(rather than concentration in a single bucket like Scam2Prompt) is
what drives this shift: it does not saturate any single
under-represented category, but it brings each of firmware, IoT, and
supply-chain from one-corpus-with-token-coverage to two-corpus
coverage at materially higher absolute counts.
\emph{Post-compromise persistence}, while present in five corpora
(now including ASTRA), is consistently a minority bucket within
each. \emph{Cryptographic-implementation bugs} and AI-system attacks
(prompt-injection-as-malware-elicitation, agent-poisoning) are
similarly thin outside CyberSecEval~2--3. These remaining gaps mean
that current refusal benchmarks still underestimate coverage of
attacker capabilities that increasingly dominate real-world incident
reports, although less severely than at the time of the previous
review cut-off.

\subsection{Categories Plausibly Out of Scope}
\label{sec:coverage_oos}
Some apparent under-coverage is not a corpus defect but an artefact
of the elicitation modality. Single-turn natural-language code-generation
prompts are a poor fit for post-exploitation ROP-chain construction,
hardware-specific kernel rootkits, and side-channel attacks whose
exploitation hinges on micro-architectural state rather than source
code; firmware exploits typically require a target image and a debug
harness rather than a free-form prompt. These categories may be
genuinely under-elicitable in the single-turn coding-prompt format and
better measured through agentic, multi-step protocols such as MOCHA's
decomposition attacks or CyberSecEval~3's autonomous-offensive-operations
subset.  We flag the distinction explicitly so that future corpus design
does not conflate \emph{under-representation} (a coverage gap that new
prompts could close) with \emph{non-applicability} (a modality mismatch
that requires a different evaluation construct).


\section{Discussion}
\label{sec:discussion}

\subsection{Principal Findings}

Applying a uniform extraction template to thirteen publicly released
corpora (\S\ref{sec:methodology_comparison},
\S\ref{sec:reproducibility}) surfaces three findings that are striking
less for their content than for their uniformity:

\begin{itemize}
  \item \textbf{Zero of thirteen corpora report inter-rater reliability
        on their refusal labels.} No corpus reports a Fleiss-$\kappa$
        statistic, or any analogous per-prompt agreement coefficient
        between independent annotators, on its released refusal
        labels.
  \item \textbf{Zero of thirteen corpora gate access at first release.}
        Every artefact ships open-access at preprint stage, even when
        the artefact contains executable malware payloads or
        adversarial templates with documented jailbreak success rates
        above 60\,\%.
  \item \textbf{Zero of thirteen corpora document a takedown policy.}
        No primary publication or release artefact names a
        responsible-disclosure contact, content-removal pathway, or
        takedown timeline.
\end{itemize}

The qualifier \emph{released refusal labels} matters for the first
finding: several corpora derive labels from automated oracles
(CySecBench, RedCode, CIRCLE, Scam2Prompt) or from upstream-repository
provenance (MCGMark), and JailbreakBench ships a 300-item
human-preference calibration set for its judge. ASTRA distinguishes
itself among the in-scope corpora by shipping a per-prompt
\texttt{malicious\_rationale} annotation alongside each request,
explicitly characterising the harmful intent that motivates the prompt;
it is the only in-scope corpus that does so. What remains absent
across the thirteen entries is a per-prompt agreement statistic between
independent annotators on the released refusal labels themselves.

None of these absences is remarkable in isolation; an individual paper
might reasonably defer reliability reporting, gating, or takedown
documentation to a follow-up release. The pattern is remarkable
because it is uniform across thirteen independently authored corpora
spanning three years and at least seven research groups, which
suggests that the methodological scaffolding for refusal-corpus
construction has not yet stabilised into a shared norm. The
cross-corpus methodology comparison
(Table~\ref{tab:corpus_methodology}) and the reproducibility table
(Table~\ref{tab:licensing}) make these uniform absences visible in a
way that no single corpus paper can.

%

\begin{figure}[!htbp]
  \centering
  \includegraphics[width=\linewidth]{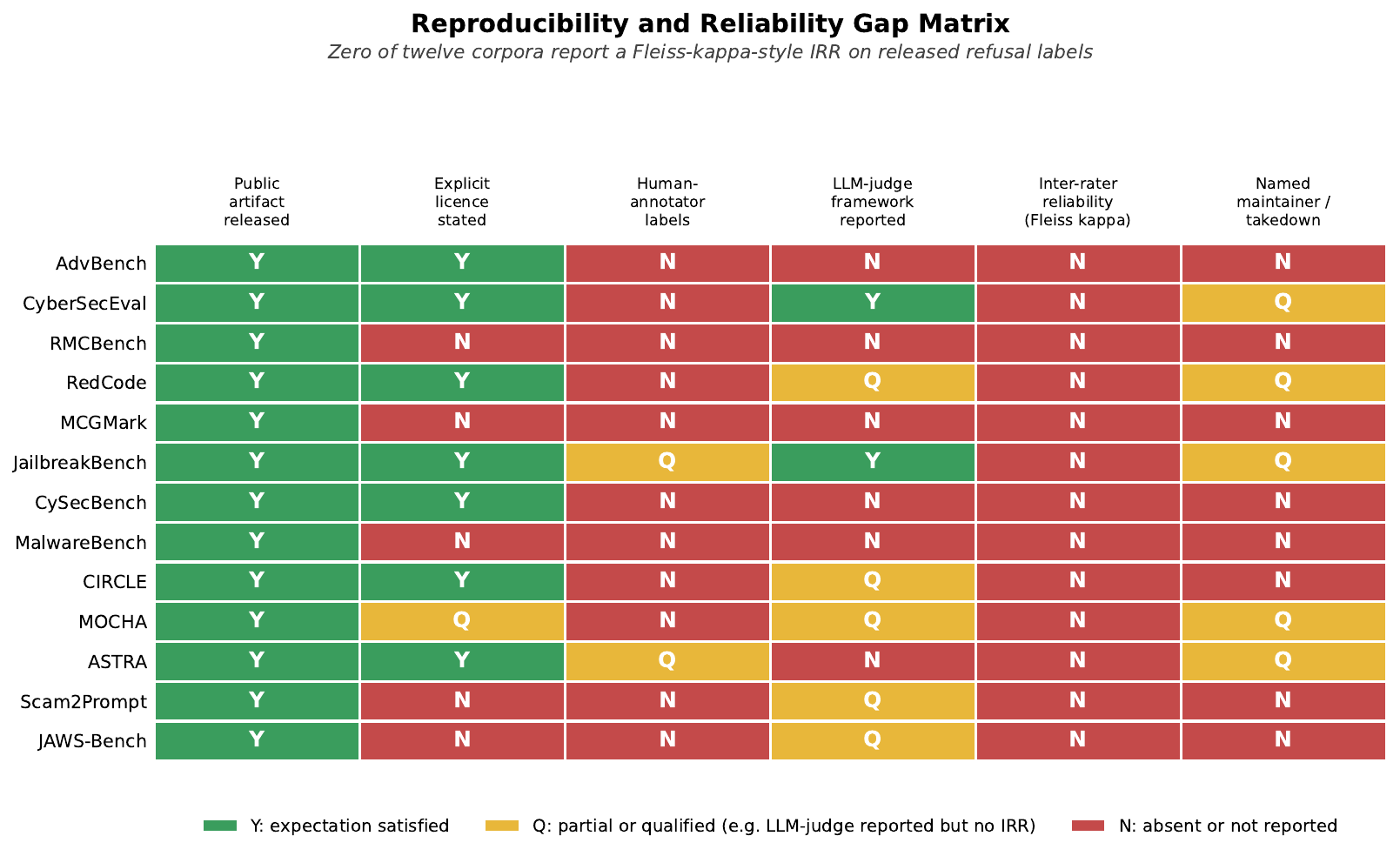}
  \caption[Reproducibility and Reliability Gap Matrix]{Reproducibility
  and Reliability Gap Matrix for the thirteen in-scope
  malicious-code-generation prompt corpora. Rows are corpora in
  chronological release order; columns are six methodological
  expectations consolidated from
  Tables~\ref{tab:corpus_methodology} and~\ref{tab:licensing}.
  Each cell is one of three states: green Y (expectation
  satisfied), amber Q (partial or qualified, e.g.\ an LLM-judge
  framework is reported but no Fleiss-$\kappa$ inter-rater
  reliability is computed), or red N (expectation absent or not
  reported). The IRR column is uniformly red: zero of thirteen
  surveyed corpora report a Fleiss-$\kappa$-style inter-rater
  reliability statistic on their released refusal labels.}
  \label{fig:gap-matrix}
\end{figure}

Figure~\ref{fig:gap-matrix} visualises the cross-corpus methodology
audit on six axes: artefact release, explicit licence, human-annotator
labels, LLM-judge framework, inter-rater reliability on released
refusal labels, and a named maintainer or takedown contact. The first
column is uniformly green (every in-scope corpus is publicly released),
and the licence column is mixed: roughly half the corpora state an
explicit licence, the rest leave reuse terms unstated at preprint
stage. The two reliability columns tell the headline story. Only
JailbreakBench's 300-item human-preference calibration set qualifies
as a partial human-annotator anchor (cell coloured amber), and the
inter-rater reliability column is red across all thirteen rows. The
maintenance column is comparably sparse: only four corpora host an
active issues channel that approximates a takedown contact, and none
publish a named-maintainer governance statement. The matrix replaces
roughly one paragraph of prose in
\S\ref{sec:methodology_comparison} and \S\ref{sec:reproducibility}
and supports the recommendations in
\S\ref{sec:discussion}\cite{verga2024juries,landis1977measurement}.

\subsection{Open Methodological Problems}

Three problems recur across the surveyed corpora. The
absence-of-human-baselines critique below concerns interpretive
labeling of individual refusal decisions on individual prompts, not
the descriptive cataloging task this review itself undertakes
(\S\ref{sec:method_screen}); the two are distinct reliability tasks
operating on distinct objects, and the present review does the
descriptive task (transparently, every value traceable to a source)
while critiquing the surveyed literature for the interpretive task
(which is uniformly absent).

\textbf{Absence of human-annotator baselines.} No surveyed corpus
reports human-labeller agreement against which LLM-judge labels could
be calibrated. Where corpora rely on automated classifiers or
execution-trace scoring (CySecBench, MOCHA, RedCode, CIRCLE), the
reported refusal-rate statistics inherit whatever bias the judging
model carries, with no human anchor against which that bias can be
measured. Movva et al.\ \cite{movva2024annotation} report a
GPT-4-to-human Pearson correlation of only $r = 0.59$ on conversational
safety annotation, which means that an LLM judge agreeing strongly with
itself across runs can still diverge substantially from a human rater
on the underlying construct. Without a human baseline, refusal-rate
statistics computed by LLM judges sit on an uncalibrated foundation,
and downstream comparisons across corpora compound the uncertainty
silently.

\textbf{Lack of agreed-upon prompt-composition standards.} Cross-corpus
refusal-rate comparisons are confounded by differences in
prompt-distribution composition. RMCBench's 473 prompts mix text-to-code
with code-to-code variants; MalwareBench's 3{,}520 prompts mix 320 bare
malware seeds with 11 jailbreak-decorated variants per seed;
CySecBench's 12{,}662 prompts are LLM-generated paraphrases over ten
attack types. A model that refuses 80\,\% of MalwareBench and 80\,\% of
CySecBench is not necessarily refusing the same construct twice. The
field needs an explicit composition manifest (modality split, jailbreak
vs.\ bare-prompt split, seed vs.\ paraphrase split) before headline
refusal rates can be treated as commensurable.

\textbf{Fragmentation of malware-category taxonomies.} No canonical
schema spans all thirteen corpora. The union ontology in
\S\ref{sec:coverage} (22 categories) was constructed for this review and
is not a community standard. The CyberSecEval family uses MITRE-derived
categories, MOCHA uses 13 functional families, CySecBench uses 10 attack
types, and the rest either use implicit clustering or no formal
taxonomy at all. None of these schemas cleanly maps onto another, which
forces every cross-corpus analysis (including this one) to perform an
ad-hoc reconciliation step before comparison is possible.

\subsection{Recommendations for Future Corpus Construction}

The following six items are proposed methodological directions for
next-generation corpora; they are not standards established by this
review. The synthesis above establishes that current practice is
uneven on each dimension below, but the specific direction proposed
here is a community judgment, not an empirical finding of this paper.
Recommendations 2, 3, and 6 have a concrete implementation in the
companion prompt-bank work \cite{young2026promptbank}, which we cite
as an exemplar of the pattern rather than as the only way to satisfy
each recommendation. Each item is tied to a specific gap surfaced by
the extraction template in
\S\ref{sec:methodology_comparison}--\S\ref{sec:reproducibility}.

\begin{enumerate}
  \item \textbf{Pre-register inclusion criteria} before construction
        begins, so that the threat-model boundary is documented
        prospectively rather than reconstructed by readers from the
        artefact description. This addresses the inclusion-criteria
        fragmentation observed in \S\ref{sec:methodology}.
  \item \textbf{Use a vendor-diverse multi-judge panel for label
        validation.} The jury-based reliability framework proposed by
        Verga et al.\ \cite{verga2024juries} is the natural starting
        point; the five-judge implementation in
        \cite{young2026promptbank} demonstrates one concrete realisation
        spanning four vendor families. This addresses the
        human-baseline gap discussed in \S\ref{sec:discussion}.
  \item \textbf{Report Fleiss' $\kappa$ with bootstrap 95\,\% CI as a
        baseline expectation} alongside any refusal-rate statistic, as
        per the Landis and Koch \cite{landis1977measurement} convention.
        The companion prompt-bank work \cite{young2026promptbank}
        implements this recommendation, reporting Fleiss' $\kappa$ with
        bootstrap 95\,\% CI across its five-judge panel on the prompts
        on which all five judges returned a valid label, both overall
        and per source dataset. This addresses the IRR-equals-zero
        finding in \S\ref{sec:methodology_comparison}.
  \item \textbf{Adopt a consistent malware-category taxonomy.} We
        propose the 22-category union ontology in
        \S\ref{sec:coverage} as a candidate community standard, to be
        debated and revised rather than imposed; the value of any such
        ontology is in shared adoption, not in any particular author's
        choice of granularity. This addresses the taxonomy
        fragmentation problem.
  \item \textbf{Document a license, takedown, and gating policy at
        release}, even if the policy is minimal. A two-line statement
        on the arXiv landing page (license, named maintainer email,
        takedown timeline) would make the governance posture of the
        release auditable in a way that current practice does not, and
        would close the gated-access-equals-zero and
        takedown-equals-zero findings in
        \S\ref{sec:reproducibility} at the level of audit trail
        (whether such a policy is the right policy, or whether the
        gating threshold should be set high or low, is a further
        question this review does not adjudicate).
  \item \textbf{Pair every released prompt corpus with at least one
        downstream behavioural benchmark} on a target-model panel. The
        prompt corpus is the substrate; the behavioural evaluation is
        what makes it useful. The companion prompt-bank work
        \cite{young2026promptbank} together with an unpublished
        thirteen-model behavioural baseline by the present authors is
        one concrete realisation of this pairing: the labeled prompt
        bank is paired with a behavioural panel rather than released
        standalone. Several surveyed corpora ship without an
        accompanying behavioural measurement, leaving downstream
        researchers to reconstruct the panel themselves and re-introducing
        the cross-study composition variance that the rest of these
        recommendations are intended to remove.
\end{enumerate}

These recommendations are deliberately incremental: each tracks a
specific gap surfaced by the extraction template
(\S\ref{sec:methodology_comparison}, \S\ref{sec:reproducibility}) and
none requires a methodological breakthrough. The remaining barrier is
coordination, not technique.

\subsection{Limitations of This Review}

Four limitations qualify the synthesis above.

\textbf{Author conflict-of-interest disclosure.} Two of the
methodological substrates referenced in this review are companion
works from the same research group: the prompt-bank consolidation
\cite{young2026promptbank} and an unpublished thirteen-model
behavioural baseline by the present authors. References to these
works are confined to points where they uniquely contribute (the
five-judge framework, the behavioural-benchmark pairing
recommendation), and they appear as ordinary entries in the
synthesis rather than as the framing of the review.

\textbf{Search cutoff.} The search cutoff is the final search-execution date, 2026-05-06; corpora
released after that date are out of scope until a planned update pass before submission. The field is moving rapidly
enough that a 12--18 month update is plausible and indeed planned (see
\S\ref{sec:future_work}).

\textbf{No re-measurement.} This review is descriptive: it reports each
corpus as released, without re-measuring inter-rater reliability under a
uniform protocol. A uniform IRR re-evaluation is identified as future
work in \S\ref{sec:future_work} rather than attempted here.

\textbf{Single-screener pass on Tier-3 exclusions.} Tier-1 and Tier-2
candidates were screened by both authors with disagreement resolved
through discussion. Tier-3 confirmed exclusions (corpora cited only as
adjacent context) were screened by one author, with the second author
auditing the Tier-3 list as a whole rather than each excluded entry
individually.

\subsection{Future Work}
\label{sec:future_work}

The most direct extension of this review is a planned empirical
re-evaluation. The follow-on study \cite{young2026paper5} would apply
a uniform multi-judge consensus protocol, namely the five-judge panel
from \cite{young2026promptbank}, to all thirteen in-scope corpora,
either in full (where corpus size permits) or on a stratified
sub-sample of approximately 300 prompts per corpus; report Fleiss'
$\kappa$ with bootstrap 95\,\% confidence intervals on directly
comparable terms; and produce a forest plot of original-versus-re-measured
$\kappa$ per corpus. Such a re-evaluation would convert the
descriptive catalogue produced here into an empirically anchored
cross-corpus reliability study.
Two further directions follow from the gaps identified in
\S\ref{sec:coverage}: construction of a corpus targeting the
under-represented categories surfaced by the coverage map
(supply-chain, firmware, post-compromise persistence), and a periodic
``living survey'' update on a 12--18 month cadence to track corpus
releases past the 2026-05-06 search-execution cutoff. The Paper~1 prompt-bank
\cite{young2026promptbank} and an unpublished companion thirteen-model behavioural baseline together
provide the methodological substrate
on which any future empirical re-evaluation would rest: the consensus
protocol and the target-model panel respectively. A longer-horizon
direction is the construction of a community-maintained registry of
malicious-code prompt corpora that absorbs each new release into the
extraction template proposed here, so that the cross-corpus comparison
visible in Tables~\ref{tab:corpus_methodology} and~\ref{tab:licensing}
is kept current rather than reconstructed from scratch every two years.

\section{Conclusion}
\label{sec:conclusion}

This paper presents the first PRISMA-style systematic review that
treats malicious-code-generation prompt corpora as the unit of
analysis, rather than as adjacent context to a broader code-security
or jailbreak-attack taxonomy. The protocol combined keyword search
across five engines with backward-citation snowballing and full-text
screening of borderline candidates; thirteen corpora were retained,
and a uniform extraction template was applied to each. The synthesis yields four
artefacts: a cross-corpus methodology comparison
(\S\ref{sec:methodology_comparison}), a prompt-construction taxonomy
(\S\ref{sec:taxonomy}), a reproducibility and licensing landscape
(\S\ref{sec:reproducibility}), and a malware-category coverage map
(\S\ref{sec:coverage}). The headline finding is that across the
thirteen corpora, three methodological features are uniformly absent:
Fleiss-$\kappa$-style inter-rater reliability reporting, gated-access
release policies, and documented takedown commitments.

Two downstream effects follow. First, researchers selecting a corpus
for evaluation work can now condition that choice on construction
methodology, modality (single-turn, multi-turn, agent, interpreter),
license terms, and category coverage rather than on availability
alone, with the comparison tables serving as the audit substrate.
Second, future corpus authors inherit a checklist of methodological
expectations: the six recommendations distilled in
\S\ref{sec:discussion} cover pre-registration of inclusion criteria,
multi-judge consensus labelling with $\kappa$ reporting, explicit
licensing on the landing page, named-maintainer takedown policy,
gated access for high-malware-density releases, and stable schema
versioning. The 22-row union ontology proposed in
\S\ref{sec:coverage} is offered as a candidate community standard for
malware-category taxonomies, intended to replace the per-paper
taxonomies that currently prevent direct cross-corpus aggregation.
Taken together, the review converts an ad hoc citation graph into an
auditable substrate for downstream choice and future construction.

Three directions for future work follow directly. First, the most
immediate extension is an empirical re-evaluation that applies a uniform multi-judge consensus protocol to
all thirteen corpora and reports Fleiss-$\kappa$ values directly
comparable across them, using the prompt-bank substrate
\cite{young2026promptbank} and the thirteen-model behavioural
baseline (unpublished companion work) as starting infrastructure.
Second, the under-represented-categories finding in
\S\ref{sec:coverage} is itself a corpus-construction agenda:
supply-chain attacks, firmware exploits, post-compromise persistence,
and hardware and side-channel attacks all merit dedicated future
releases that treat these categories as primary rather than residual.
Third, the field moves rapidly enough that a living-survey update on
a 12 to 18 month cadence is warranted, with subsequent updates
anchoring incremental corpus releases against the methodological
scaffolding established here so that newcomers do not have to
reconstruct the comparison axes from scratch.

\section*{Acknowledgments}
This review draws directly on the work of the authors and maintainers
of the thirteen in-scope corpora, whose construction, labelling, and
release decisions made cross-corpus comparison possible. Building and
maintaining a malicious-code prompt corpus is technically demanding,
ethically fraught, and largely unrewarded by traditional academic
incentives. The authors also acknowledge the safety, alignment, and
red-teaming researchers whose continuous auditing of frontier coding
LLMs renders refusal a measurable construct, and the
responsible-disclosure maintainers at major model providers and across
the open-weights community whose triage of jailbreak reports keeps the
deployed surface tractable for downstream evaluation work.

\bibliographystyle{unsrt}
\bibliography{references}

\clearpage
\appendix

\section{Search Protocol Detail}
\label{app:search}

This appendix documents the literature-search protocol used to identify the
thirteen malicious-code-generation prompt corpora reviewed in the body of the
paper. The protocol follows the PRISMA-style conventions adopted in
\S\ref{sec:methodology}, adapted for a corpus-level systematic
review whose unit of analysis is a released prompt dataset rather
than an individual study. All search artefacts (raw query results, per-batch snowball notes,
and full-text screening decisions) are preserved for reproducibility
in the supplementary search-log archive accompanying this paper.

\subsection{Search Engines and Date Cutoff}
\label{app:search:engines}

The primary keyword and snowball searches were executed on 2026-05-05,
and an extended cross-venue search (OpenReview, ACM~DL, IEEE~Xplore,
PaperswithCode, DBLP, and Semantic~Scholar) was executed on 2026-05-06.
The cutoff for inclusion is therefore the search-execution date,
2026-05-06: any preprint or proceedings paper first available after that
date is out of scope for this review and is a candidate for a planned
update pass before submission. Searches were run against the following
engines:

\begin{itemize}
  \item \textbf{arXiv API} via \texttt{export.arxiv.org/api/query}, used as
    the primary search engine. Nine query strings (Section~\ref{app:search:queries})
    were executed against the full-text and metadata indices, with category
    filters applied where noted.
  \item \textbf{Google Scholar} and direct web search, used to recover
    non-arXiv venues and to cross-check coverage. Four web queries
    (Section~\ref{app:search:queries}) targeted NeurIPS proceedings (Datasets
    and Benchmarks track), the ACM Digital Library (CCS, USENIX Security),
    and the ACL Anthology (ACL, EMNLP).
\end{itemize}

The arXiv API was treated as the system of record for citation IDs; web
searches contributed venue metadata and surfaced corpora that were either
unavailable on arXiv or not retrievable under our query strings.

\subsection{Search Queries}
\label{app:search:queries}

The nine arXiv API queries, with the number of unique hits returned by each,
are reproduced below (URL-encoded parameters decoded to readable form):

\begin{enumerate}
  \item \texttt{all:"malicious code" AND all:"language model" AND (all:benchmark OR all:dataset)} (23 hits)
  \item \texttt{all:"code safety" AND (all:refusal OR all:jailbreak) AND all:"language model"} (0 hits, phrase tokenization issue noted)
  \item \texttt{all:"harmful code" AND all:LLM AND (all:benchmark OR all:evaluation)} (4 hits)
  \item \texttt{all:jailbreak AND all:code AND all:refusal AND (all:benchmark OR all:dataset)} (30 hits)
  \item \texttt{all:cybersecurity AND all:LLM AND all:prompt AND ti:(benchmark OR dataset)} (8 hits)
  \item \texttt{all:"coding LLM" AND all:safety AND (all:benchmark OR all:dataset)} (10 hits)
  \item \texttt{ti:CyberSecEval OR ti:HarmBench OR ti:WMDP OR ti:SecRepoBench OR ti:SafeCoder} (8 hits)
  \item \texttt{all:agent AND all:code AND all:safety AND (all:benchmark OR all:dataset) AND (cat:cs.CR OR cat:cs.SE)} (30 hits)
  \item \texttt{all:"prompt dataset" AND all:LLM AND (all:safety OR all:harmful)} (12 hits)
\end{enumerate}

After deduplication across the nine queries, approximately 120 unique arXiv
identifiers were surfaced. The four web queries supplementing the arXiv
search were:

\begin{enumerate}
  \item RealSec-bench Wang 2026 LLM code security benchmark Java repository (Google)
  \item USENIX Security CCS 2024--2026 LLM malicious code refusal benchmark dataset (Google)
  \item NeurIPS Datasets Benchmarks track 2024--2025 LLM code safety malicious refusal evaluation (Google)
  \item ACL EMNLP 2025 coding LLM safety benchmark refusal malware prompts (Google)
\end{enumerate}

\subsection{Snowball Backward-Citation Protocol}
\label{app:search:snowball}

To guard against gaps in keyword-based retrieval, a backward-citation snowball
was performed on every in-scope corpus paper. For each paper, the
Related Work section and the full reference list were scanned for additional
candidate corpora that match the inclusion criteria (released prompt dataset,
malicious-code-elicitation, executable-code modality). Citations to attack
methods, vulnerability-detection benchmarks, and general-safety preference
datasets were noted but not promoted into the candidate pool.

The snowball was divided into five screening batches, each covering a set of
in-scope papers (CyberSecEval 1, 2, and 3 were treated as three separate
papers and batched together; the other corpora were grouped to balance batch
size). Across the five batches, approximately 190 references were scanned in
total. Per-batch outputs are preserved in the \texttt{snowball/}
subdirectory of the supplementary search-log archive.

\subsection{Screening}
\label{app:search:screening}

Candidate papers passed through a two-pass screen. The first pass applied
the inclusion criteria of Section~3.2 to the title and abstract of each
candidate; papers that were clearly out of scope (multimodal-only,
inverse-task vulnerability avoidance, knowledge-elicitation MCQ, attack
methods with no released corpus) were excluded at this stage with the
exclusion reason recorded. The second pass was a full-text screen of
borderline candidates, in which the methods section, the released artefact
description, and any data-availability statement were read in full.

Eight borderline candidates surfaced by the snowball underwent Tier-2
full-text screening. The screen yielded \textbf{one promotion} (MCGMark /
MCGTest, arXiv 2408.01354 \cite{ning2024mcgmark}) and seven confirmed
exclusions. The extended cross-venue search added two further
promotions: Scam2Prompt / Innoc2Scam-bench
\cite{chen2025scam2prompt} and JAWS-Bench
\cite{saha2025jawsbench}. A third addition, ASTRA / astra-agent-security
\cite{xu2025astra}, arrived through a separate identification path:
on 2026-05-10, after the database, snowball, and extended cross-venue
searches were complete, the lead author of the ASTRA paper emailed
the present authors and brought the corpus and its accompanying
Hugging Face dataset (\texttt{PurCL/astra-agent-security}, MIT licence,
1{,}995 prompts) to our attention. The same three inclusion criteria
(\S\ref{sec:method_criteria}) plus the exclusion screen were applied
to ASTRA as to the other twelve corpora; all three were met and the
exclusion screen was cleared, and the corpus was added to the
in-scope list. The author-contact identification path is recorded
here as a distinct provenance category rather than back-fitted into
the search-engine batches, so the audit trail of how each in-scope
corpus entered the review remains transparent. As a provenance note for reviewers, JAWS-Bench
was desk-rejected at ICLR~2026 (OpenReview submission
ID~\texttt{dpVhlf13tk}); the present review treats it as the latest
publicly-available preprint version (arXiv:2510.01359~v1, 2025-10-01),
which is consistent with the inclusion criterion that the artefact be
publicly described in a primary publication. The seven Tier-2
exclusions from the snowball-derived borderline screen are:

\begin{itemize}
  \item \textbf{Jailbroken} \cite{wei2023jailbroken}: a methodological
    analysis of jailbreak failure modes; no released refusal-evaluation
    corpus.
  \item \textbf{BeaverTails} \cite{ji2023beavertails}: a generic
    safety-preference dataset for RLHF, not coding-specific; the
    malware/hacking sub-extract is general harm rather than executable-code
    requests.
  \item \textbf{h4rm3l} \cite{doumbouya2024h4rm3l}: a domain-specific
    language for composing jailbreak attacks; underlying intents are
    general harm, not coding-specific, and the framework is orthogonal to
    any of the in-scope corpora.
  \item \textbf{SecCodePLT} \cite{nie2024seccodeplt}: an inverse-task
    benchmark covering secure-coding, vulnerability-detection, and patch
    generation, already excluded under the vulnerability-avoidance cluster.
  \item \textbf{AgentHarm} \cite{andriushchenko2024agentharm}: a multi-step
    agent-misuse benchmark whose cybercrime sub-extract (approximately ten
    prompts) is too small to constitute a comparable unit of analysis.
  \item \textbf{HackAPrompt} \cite{schulhoff2023hackaprompt}: a
    600K-prompt corpus of crowdsourced prompt-injection attacks; the
    threat model (hijacking an application's instruction template) is
    distinct from the malicious-code-refusal threat model considered here.
  \item \textbf{LLMSmith} \cite{liu2024llmsmith}: an enumeration of RCE
    and SQLi vulnerabilities in eleven LLM-integrated frameworks; the
    artefact is a list of CVEs and exploits against named LLM applications,
    not a refusal-evaluation prompt corpus.
\end{itemize}

\subsection{PRISMA Flow Statement}
\label{app:search:prisma}

\begin{figure}[htbp]
  \centering
  \includegraphics[width=0.78\linewidth]{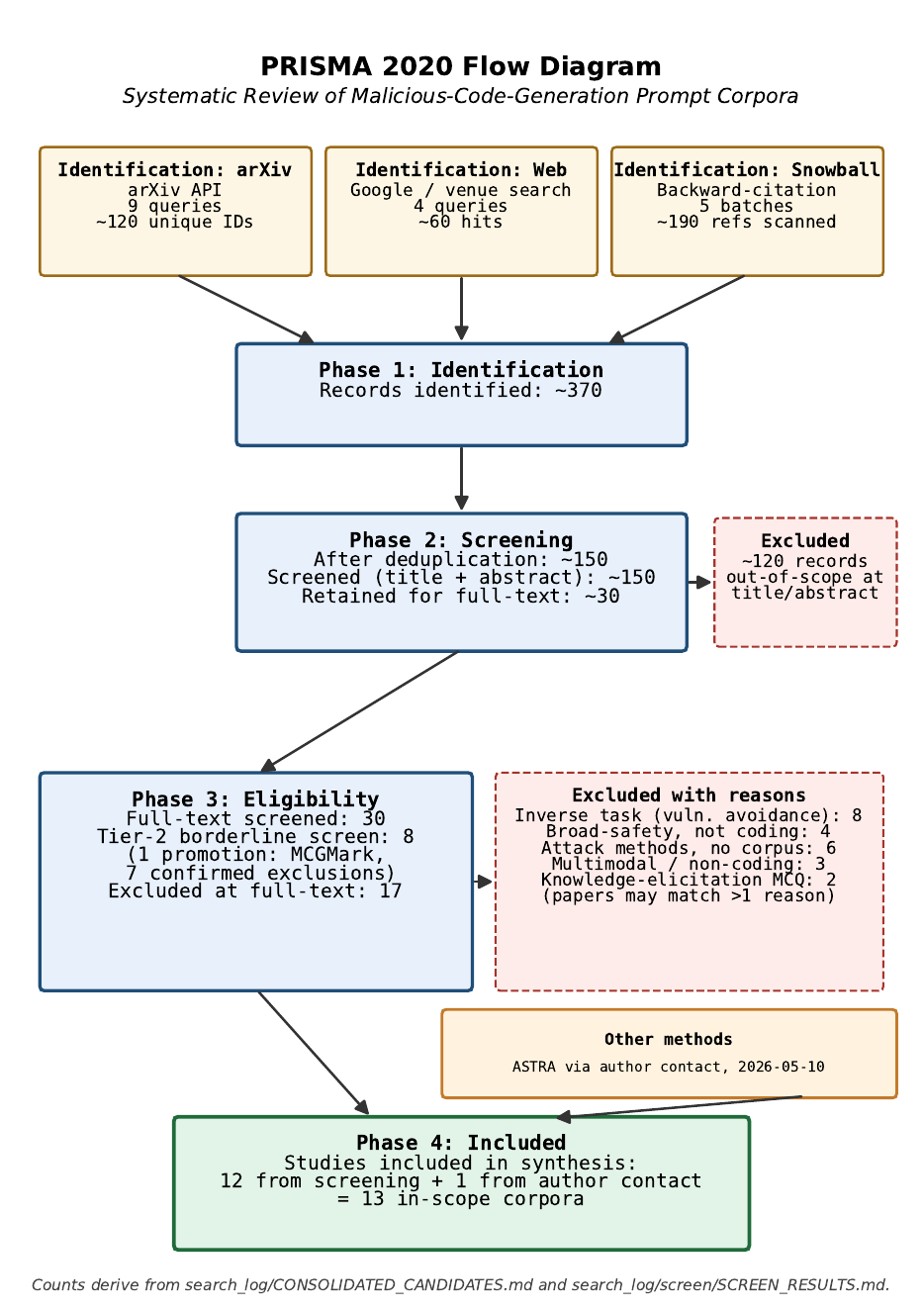}
  \caption{%
    PRISMA 2020 flow diagram for the systematic review of
    malicious-code-generation prompt corpora. Phase 1 (identification)
    aggregates approximately 370 records and reference items, comprising
    $\sim$200 database/web records (arXiv API, Google Scholar, NeurIPS
    proceedings, ACL Anthology, USENIX/ACM Digital Libraries) and
    $\sim$190 references inspected via backward-citation snowballing.
    Phase 2 (screening) deduplicates to roughly 150 records and excludes
    about 120 at the title/abstract stage. Phase 3 (eligibility) full-text screens 30
    candidates and excludes 17 with reasons. Phase 4 retains 13 in-scope
    corpora for synthesis: 9 from the primary arXiv plus web search, plus
    1 promoted from a Tier-2 borderline screen (MCGMark / MCGTest), plus
    2 promoted from the extended OpenReview and ACM~DL search
    (Scam2Prompt / Innoc2Scam-bench and JAWS-Bench), plus 1 added on
    2026-05-10 via direct author contact (ASTRA / astra-agent-security,
    Xu et al.\ 2025). The author-contact path is documented as a
    distinct identification source in \S\ref{app:search:snowball}.%
  }
  \label{fig:prisma-flow}
\end{figure}

We report the PRISMA flow both as a textual statement and as the diagram
in Figure~\ref{fig:prisma-flow}:

\begin{itemize}
  \item \textbf{Records identified.} Approximately 200 records were
    identified across arXiv (nine queries, $\sim$120 unique IDs after
    deduplication), Google Scholar, NeurIPS proceedings (Datasets and
    Benchmarks track), the ACL Anthology, and the USENIX/ACM Digital
    Libraries via direct database/web searches. A further $\sim$190
    references were inspected via backward-citation snowballing of the
    in-scope corpus papers, yielding a combined identification-stage
    count of approximately 370 records and reference items before
    deduplication, the figure shown in the PRISMA diagram
    (Figure~\ref{fig:prisma-flow}).
  \item \textbf{After deduplication.} Approximately 150 unique records
    remained after removing cross-engine duplicates and arXiv-versus-venue
    duplicates of the same artefact.
  \item \textbf{Title and abstract screen.} 30 candidates were retained:
    the thirteen in-scope corpora, eight Tier-2 borderline candidates, and
    approximately nine adjacent surveys and methodology papers cited in
    Sections~2 and 6.
  \item \textbf{Full-text screen.} Thirteen in-scope corpora were retained:
    nine from the primary arXiv plus web search, one promoted from the
    Tier-2 borderline screen (MCGMark / MCGTest), two promoted from
    the extended OpenReview / ACM~DL search (Scam2Prompt /
    Innoc2Scam-bench and JAWS-Bench, the latter of which is a
    code-agent jailbreaking benchmark with three workspace regimes
    and an executable-aware judge framework), and one added on
    2026-05-10 via direct author contact (ASTRA /
    astra-agent-security). One
    corpus that appeared in earlier drafts of the in-scope list
    (SecureAgentBench / SecureVibeBench) was reclassified to the Tier-3
    inverse-task exclusion cluster on full-text review;
    see \S\ref{sec:exclusions}.
  \item \textbf{Reasons for exclusion at the full-text stage.} Of the
    candidates excluded at full-text screening, with papers allowed to
    match more than one reason code: 8 inverse-task
    (vulnerability avoidance and secure-code generation), 4 broad-safety
    (not coding-specific), 6 attack-method (no released corpus),
    3 multimodal or non-coding, 2 knowledge-elicitation MCQ, with the
    remainder methods or method-only papers without a released artefact.
\end{itemize}

\subsection{Saturation Argument}
\label{app:search:saturation}

The saturation criterion for closing the search was an already-in-list hit
rate sufficient to indicate that the connected component of the
malicious-code-refusal-corpus literature had been mostly recovered.
Backward-citation snowballing of all in-scope papers' Related Work and
References yielded \textbf{24 already-in-list hits across approximately 190
references scanned}, with no new candidate corpora surfaced that survived
the inclusion criteria as Tier-1 promotions. The Tier-2 screen of the eight
borderline candidates added a single corpus (MCGMark / MCGTest), and the
extended cross-venue search added two further pre-cutoff corpora
(Scam2Prompt / Innoc2Scam-bench and JAWS-Bench). A thirteenth entry,
ASTRA / astra-agent-security, was added on 2026-05-10 via direct
author contact and is documented as a distinct identification source
in \S\ref{app:search:snowball}, bringing the final list to thirteen entries. The
combination of (i) a high already-in-list hit rate
during the snowball, (ii) zero direct Tier-1 promotions from the snowball,
(iii) one promotion from the Tier-2 borderline screen, (iv) two
promotions from the documented extended search, and (v) one
author-contact addition is taken as evidence that the closed search
protocol reached saturation at the time of writing. We do
not claim exhaustiveness for the post-cutoff literature: the 2026-05-06
search-execution cutoff is enforced strictly, and follow-on work should
re-execute the protocol against any later release window.

%

\end{document}